\documentclass[apj]{emulateapj}
\usepackage{epstopdf}
\usepackage{longtable}
\newcommand{\nev}{[\ion{Ne}{5}]$\lambda3426$}
\newcommand{\nevs}{[\ion{Ne}{5}]}
\newcommand{\oii}{[\ion{O}{2}]$\lambda3727$}
\newcommand{\neiii}{[\ion{Ne}{3}]$\lambda3869$}
\newcommand{\neiiis}{[\ion{Ne}{3}]}
\newcommand{\oiii}{[\ion{O}{3}]$\lambda5007$}
\newcommand{\oiiis}{[\ion{O}{3}]}
\newcommand{\feii}{\ion{Fe}{2}}
\newcommand{\mgii}{\ion{Mg}{2}$\lambda2800$}
\newcommand{\civ}{\ion{C}{4}$\lambda1549$}
\newcommand{\ciii}{\ion{C}{3}]$\lambda1909$}
\newcommand{\dv}{$\Delta V$}

\shorttitle{Outflows and Candidate Dual AGN at $z=0.8-1.6$}
\shortauthors{Barrows et al.}

\begin{document}
\submitted{accepted by ApJ}
\title{Identification of Outflows and Candidate Dual Active Galactic Nuclei in SDSS Quasars at $z=0.8-1.6$}
\author{R. Scott Barrows,\altaffilmark{1} Claud H. Sandberg Lacy,\altaffilmark{1,2} Julia Kennefick,\altaffilmark{1,2} Julia M. Comerford,\altaffilmark{3} Daniel Kennefick,\altaffilmark{1,2} and Joel C. Berrier\altaffilmark{1,2}}

\altaffiltext{1}{Arkansas Center for Space and Planetary Sciences, University of Arkansas, Fayetteville, AR 72701\newline \indent \emph{author email: rbarrows@uark.edu}}
\altaffiltext{2}{Physics Department, University of Arkansas, Fayetteville, AR 72701}
\altaffiltext{3}{Astronomy Department, University of Texas at Austin, Austin, TX 78712}
\bibliographystyle{apj}

\begin{abstract}
We present a sample of 131 quasars from the Sloan Digital Sky Survey at redshifts $0.8<z<1.6$ with double peaks in either of the high-ionization narrow emission lines  \nev~or \neiii.  These sources were selected with the intention of identifying high-redshift analogs of the $z<0.8$ active galactic nuclei (AGN) with double-peaked \oiii~lines, which might represent AGN outflows or dual AGN.  Lines of high-ionization potential are believed to originate in the inner, highly photoionized portion of the narrow line region (NLR), and we exploit this assumption to investigate the possible kinematic origins of the double-peaked lines.  For comparison, we measure the \nev~and \neiii~double peaks in low-redshift ($z<0.8$) \oiiis-selected sources.  We find that \nev~and \neiii~show a correlation between line-splitting and line-width similar to that of \oiii~in other studies; and the velocity-splittings are correlated with the quasar Eddington ratio.  These results suggest an outflow origin for at least a subset of the double-peaks, allowing us to study the high-ionization gas kinematics around quasars.  However, we find that a non-neligible fraction of our sample show no evidence for an ionization stratification.  For these sources, the outflow scenario is less compelling, leaving the dual AGN scenario as a viable possibility.  Finally, we find that our sample shows an anti-correlation between the velocity-offset ratio and luminosity ratio of the components, which is a potential dynamical argument for the presence of dual AGN.  Therefore, this study serves as a first attempt at extending the selection of candidate dual AGN to higher redshifts.
\end{abstract}


\section{Introduction}
\label{intro}
Mergers of gas-rich galaxies are likely to play a key role in the growth of supermassive black holes (SMBHs) through accretion, particularly in the triggering of quasar phases \citep{Sanders:1988,Treister:2010}.  Quasars may also represent an important phase in the evolution of galaxies due to radiative feedback \citep{Silk:Rees:1998,Kauffman:2000}.  This sequence of events is a potential scenario linking the evolution of galaxies and the growth of SMBHs \citep{Di_Matteo:2005,Hopkins05}.  Interestingly, both the merger and quasar phases can manifest themselves as active galactic nuclei (AGN) with double-peaked narrow emission lines in their spectra.  Specifically, during a galaxy merger, when the SMBHs are separated by $\sim$1 kpc and are actively accreting as a dual AGN \citep{Comerford2009a} they will each produce Doppler-shifted emission from their narrow line regions (NLRs), resulting in a double-peaked profile in the integrated spectrum.  In the feedback scenario, radiation from a single quasar can drive bi-conical outflows of the NLR \citep{Arribas:1996,Veilleux:2001,Crenshaw:2010b}, producing similar emission line profiles \citep{Zheng90,Fischer:2011}.

AGN whose spectra exhibit double-peaked narrow emission lines (i.e. they are best fit by two components) represent a sub-class of the AGN population.  Extended, offset and double-peaked line profiles have been observed in the narrow emission lines of AGN since the earliest studies of AGN NLRs \citep{Heckman1981}.  These complex narrow line profiles are observed in both Type 1 and Type 2 AGN \citep{Crenshaw:2010a}, and they are generally most pronounced in forbidden transition lines of high ionization potentials (I.P.s).  In particular, such observations have often focused on the 5007\AA~transition line of \oiiis~($I.P.=35.15$ eV) since it is a relatively intense emission line produced by the ionizing continuum of AGN and is accessible in optical spectra (see \citealt{Veilleux:1991} and \citealt{Whittle:1992} for examples).  However, there is even variation among the high-ionization lines, with those of the highest ionization potentials, such as \neiiis~($I.P.=41.07$ eV), and \nevs~($I.P.=97.16$ eV), displaying the largest velocity offsets \citep{De_Robertis:1984,Sturm:2002,Spoon:Holt:2009}.

In the dual AGN scenario, the two emission line peaks are produced by the orbital motion of two AGN within a single merger remnant galaxy.  This interpretation is intriguing since SMBHs reside in the bulges of galaxies, and dual SMBHs (kpc-scale separations) are a stage of galaxy mergers before the SMBHs coalesce.  The existence of dual AGN has been confirmed observationally in several serendipitous cases, most notably in ultra-luminous infrared galaxies as pairs of X-ray point sources \citep{Komossa2003,Guainazzi2005,Hudson2006,Bianchi2008,Piconcelli2010,Koss:2011,Mazzarella:2012}.  Systematic searches for dual AGN in large spectroscopic databases, such as the Sloan Digital Sky Survey \citep[SDSS;][]{Abazajian09} and DEEP2 \citep{Newman:2012} have involved identifying AGN with double emission line systems \citep{Comerford2009a,Wang2009,Liu2010a,Smith:2010,Ge:2012}.  Promising results are being obtained through follow-up observations in the form of high-resolution optical imaging \citep{Comerford2009b}, near-infrared (NIR) adaptive optics imaging \citep{Liu2010b,Fu:2011a,Rosario:2011,Shen:2011b,Barrows:2012}, spatially resolved spectroscopy \citep{McGurk:2011,Fu:2012}, hard X-ray observations \citep{Comerford:2011,Civano:2012,Liu:2012}, radio observations \citep{Fu:2011}, and diagnostics deduced statistically from longslit spectroscopy \citep{Comerford:2012}.  A sample of dual AGN will be useful in studying the connection between galaxy interactions/mergers and AGN activity \citep{Green2010,Liu:2011,Liu:2012a}, and for refining the galaxy merger rate \citep{Conselice:2003,Berrier:2006,Lotz:2011,Berrier:2012}. 

In the outflow scenario, offset or double-peaked narrow emission lines are often attributed to radially flowing NLR gas driven by energy from the AGN.  Furthermore, this effect is also thought to produce a stratification of the NLR since the incident ionizing flux and electron density should diminish with increasing distance from the nuclear source.  This will result in the production of different lines in varying proportions as a function of radius from the AGN \citep{Veilleux:1991}.  Observationally, luminous quasars are known to have extended NLRs driven by energy from the AGN coupled to the interstellar medium gas \citep{Bennert:2002}, and powerful radio galaxies often show complex, spatially extended NLRs with multiple components aligned along the radio axis.  Models for this alignment include reflection of the AGN emission \citep{Tadhunter:1988} or material entrained in a radio jet \citep{Holt2003,Holt2008,Komossa:2008b}.  In AGN with sufficiently high Eddington ratios, radiation pressure acting on gas and dust \citep{Everett2007b} or a hot wind that entrains the NLR clouds \citep{Everett2007a} are possible mechanisms capable of driving the line splitting.  Whichever mechanism is the dominant driver of outflows in an individual source, they may represent cases of AGN feedback, which might be important in quenching star formation in galaxies following a merger and in establishing the observed correlations between SMBH masses and host galaxy properties \citep{Ferrarese:Merritt:2000,Marconi:Hunt:2003}.

Whether AGN with double-peaked emission lines represent dual AGN or powerful AGN outflows, they have proven useful for investigating several aspects of AGN evolution, including AGN triggering and feedback.  Systematic searches for double-peaked AGN have primarily utilized the \oiii~emission line and have therefore been limited to below redshifts of $z\approx0.80$ since \oiii~is not accessible in optical spectra at higher redshifts.  However, there is evidence that at high redshifts galaxy mergers were more prevalent, and AGN outflows might have played an important role in the evolution of galaxies at $z\ge1$, including massive radio galaxies \citep{Nesvadba:2008}.  Furthermore, there is significant controversy over the role that galaxy mergers play in AGN activity and SMBH growth at high redshifts \citep{Cisternas:2011,Treister:2012,Kocevski:2012}.  Therefore, a sample of high redshift double-peaked narrow line AGN will be important for investigating these questions.  So far only one such candidate dual AGN has been identified above $z\sim0.8$ as a serendipitous discovery at $z=1.175$ through double-peaked UV and optical emission lines, particularly evident in \nev~and \neiii~\citep{Barrows:2012}.  Motivated by this discovery, we have conducted a systematic search for additional AGN at high redshift ($z>0.8$) with double-peaked \nev~and \neiii~emission lines.  These emission lines have relatively high ionization potentials and trace the gas photoionized by the AGN continuum.  Furthermore, they are likely to originate in the inner (more highly ionized) portion of the NLR, allowing for the identification of AGN driven outflows at $0.8<z<1.6$.  This study will also serve as a first attempt at extending the identification of candidate dual AGN to higher redshifts.

In Section \ref{analysis} we describe our parent sample and how we selected our final sample and low-redshift comparison sample.  In Section \ref{redshifts} we describe the systemic redshifts we will use throughout the paper.  In Section \ref{sec:offsets} we describe the general properties of our sample and our estimates of selection completeness.  In Section \ref{kinematics} we investigate several correlations among the emission line properties that aid in illuminating the origin of the double-peaked line profiles.  In Section \ref{radio} we describe the radio properties of our sample and compare them to our parent sample.  In Section \ref{interpretation} we discuss the most likely physical mechanisms driving the line-splitting, particularly focusing on the scenarios of AGN outflows and dual AGN.  In Section \ref{conclusions} we summarize our main conclusions.  Throughout the paper we adopt the cosmological parameters $\Omega_{\Lambda}=0.728$, $\Omega_{b}=0.0455$, $\Omega_{m}h^{2}=0.1347$, and $H_{0}=70.4$ km s$^{?1}$ Mpc$^{?1}$. This corresponds to the maximum likelihood cosmology from the combined WMAP+BAO+H0 results from the WMAP 7 data release of \citet{Komatsu:2011}.

\section{Generating the Sample}
\label{analysis}

\subsection{Parent Sample}
\label{sample}
Our parent sample consists of archival spectra drawn from the quasar catalog of the SDSS Data Release 7 (DR7) which is described in detail in \citet{Schneider:2010}.  The typical resolution of the SDSS spectra is $\lambda/\Delta \lambda \sim$2000.  In short, inclusion in the catalog requires luminosities brighter than $M_{i}=-22.0$, at least one emission line with FWHM $>1000$ km s$^{-1}$ or complex absorption features, apparent magnitudes fainter than $i\approx 15.0$, and have highly reliable redshifts (see Section \ref{redshifts} for a discussion of the redshifts).  We restricted the lower redshift limit of the sample to $z\ge0.80$ to only include sources \emph{not} in the parent samples of \oiiis-selected double-peaked emitters from the SDSS \citep{Wang2009,Liu2010a,Smith:2010,Ge:2012} since our intention is to select sources which are not identifiable by their methods.  Additionally, we required that at least \nev~be accessible in the SDSS wavelength range ($3800-9200$~\AA), which imposes an upper limit of $z\sim1.7$.  We did not make any selection cuts based on the signal-to-noise ratios (S/N).  This resulted in a parent sample of 39,876 sources.
~\\
\subsection{Initial Selection}
\label{sec:initial}
Our initial selection involved visually identifying quasars from the parent sample (Section \ref{sample}) with detectable double emission line peaks in \nev~and/or \neiii.  These two lines were used since they are accessible in the SDSS optical wavelength range at $z>0.80$, are relatively strong narrow lines in quasar spectra, and are not severely blended with any other strong lines.  We did not require that two explicit peaks be detectable in both of those emission lines for two reasons: 1) the ionizing continuum may be such that \nev~is too weak to be detected whereas \neiii~is detectable; and 2) a difference in line ratios and/or velocity-splittings between the two lines may result in one pair being more blended than the other.  Therefore, since the purpose of this analysis is to investigate the origin of the double-peaked emission lines, we did not want to exclude those sources which show variations among the line properties.  While \oii~is another strong emission line accessible in most of our parent sample, we did not select sources based on this line since \oii~is a doublet ($\lambda3726,3729$~\AA) with $\sim$200 km s$^{-1}$ separation between the transition wavelengths and is difficult to discern from true peaks, particularly with the limited spectral resolution of the SDSS (see \citealt{Smith:2010} for a similar discussion).  In sources where the double peak detections are ambiguous in each line, corresponding peaks in both \nev~and \neiii~are needed for confirmation.  Out of the parent sample, our visual selection process resulted in 181 preliminary sources.    \\

\begin{figure*} $
\begin{array}{c}
\hspace*{-0.1in} \includegraphics[width=7in, height=1.8in]{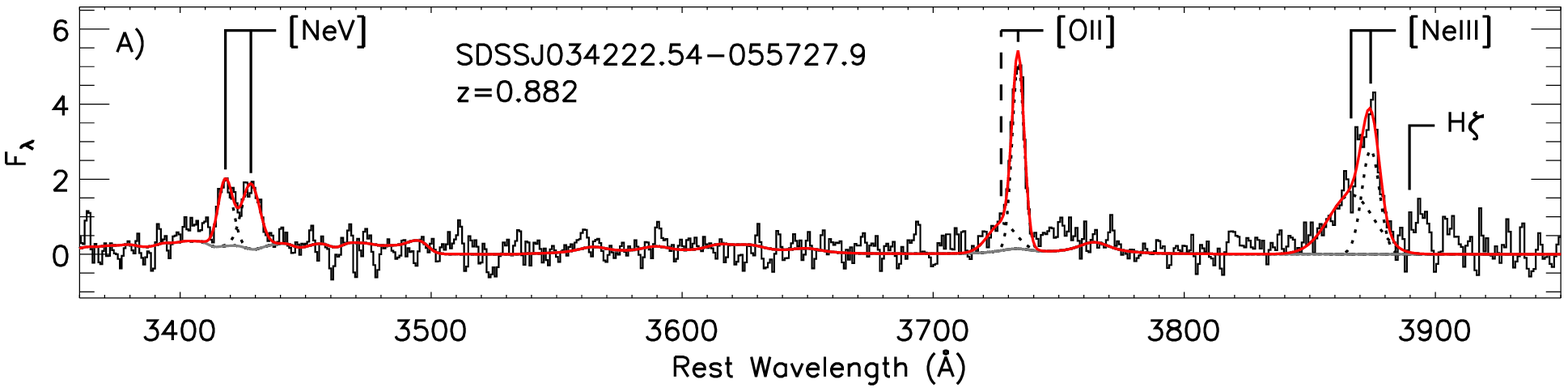} \\
\hspace*{-0.1in} \includegraphics[width=7in, height=1.8in]{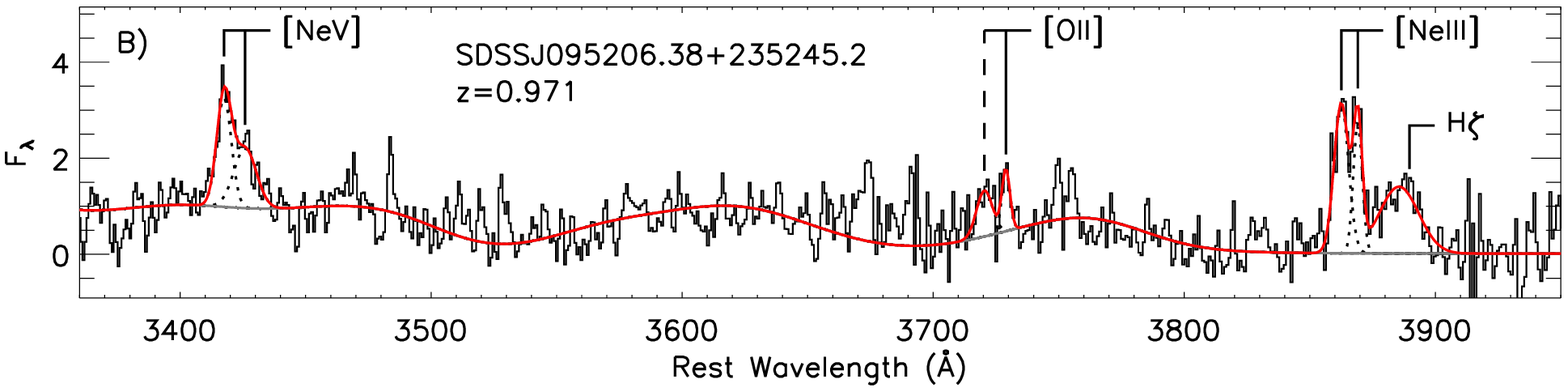} \\
\hspace*{-0.1in} \includegraphics[width=7in, height=1.8in]{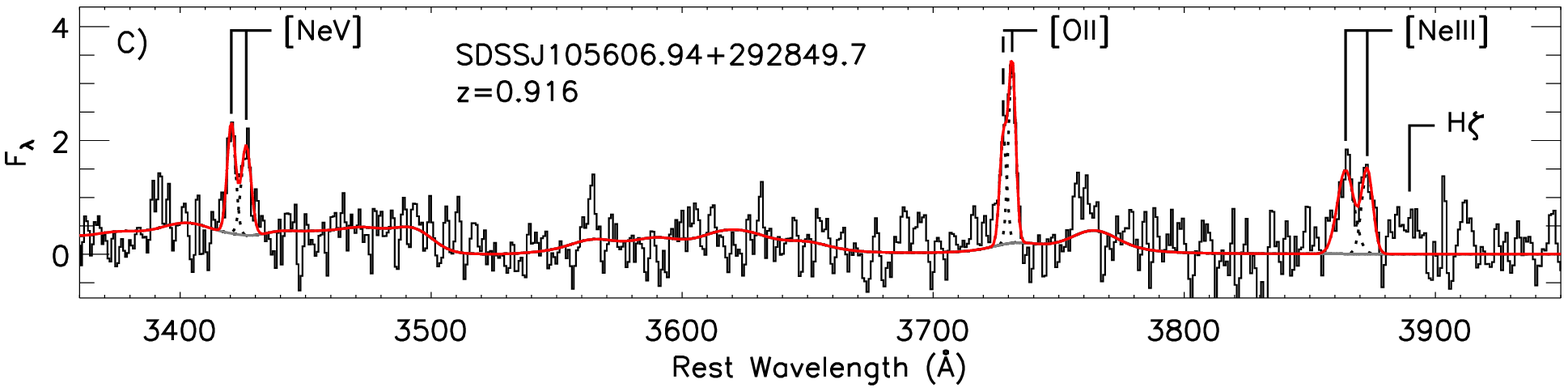} \\
\hspace*{-0.1in} \includegraphics[width=7in, height=1.8in]{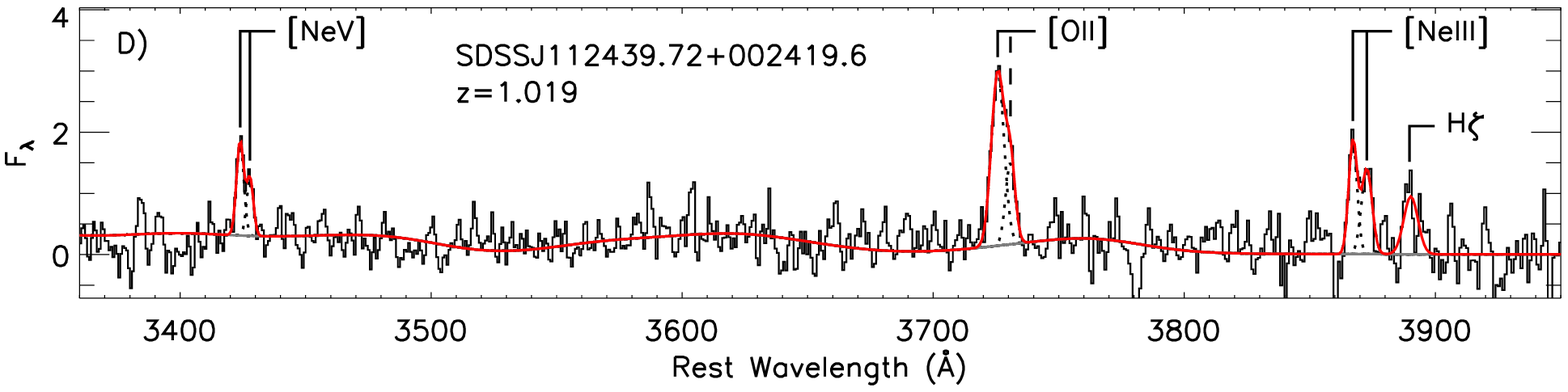}

\end{array} $
\caption{\footnotesize{Four examples of quasar spectra yielded by our selection process shown in the quasar rest-frame ($z_{SDSS}$) wavelength range of $3360-3950$~\AA.  The spectra have been smoothed by convolving with a Gaussian of $\sigma=1$~\AA.  The flux densities, $F_{\lambda}$, are in units of $10^{-17}$ erg s$^{-1}$ cm$^{-2}$~\AA$^{-1}$.  In each of the four panels the power-law continuum has been subtracted from the spectrum, and the \feii~ template (gray, solid line), best-fit Gaussians (black, dotted lines), and the best-fit model sum (red, solid line) are shown.  See Section \ref{modeling} for details on fitting the models.  These are examples for which double components are measurable in both \nev~and \neiii.  In all four cases we can only place upper limits on the presence of a second \oii~component (vertical dashed line).  In each panel we have labeled the expected position of H$\zeta$ based on its rest-wavelength, however it is not always detected, and in no cases can we resolve double-peaks in H$\zeta$.}}
\label{examples}
\end{figure*}

\begin{figure}[t!] $
\begin{array}{cc}
\hspace*{-0.2in} \includegraphics[width=1.8in, height=2.7in]{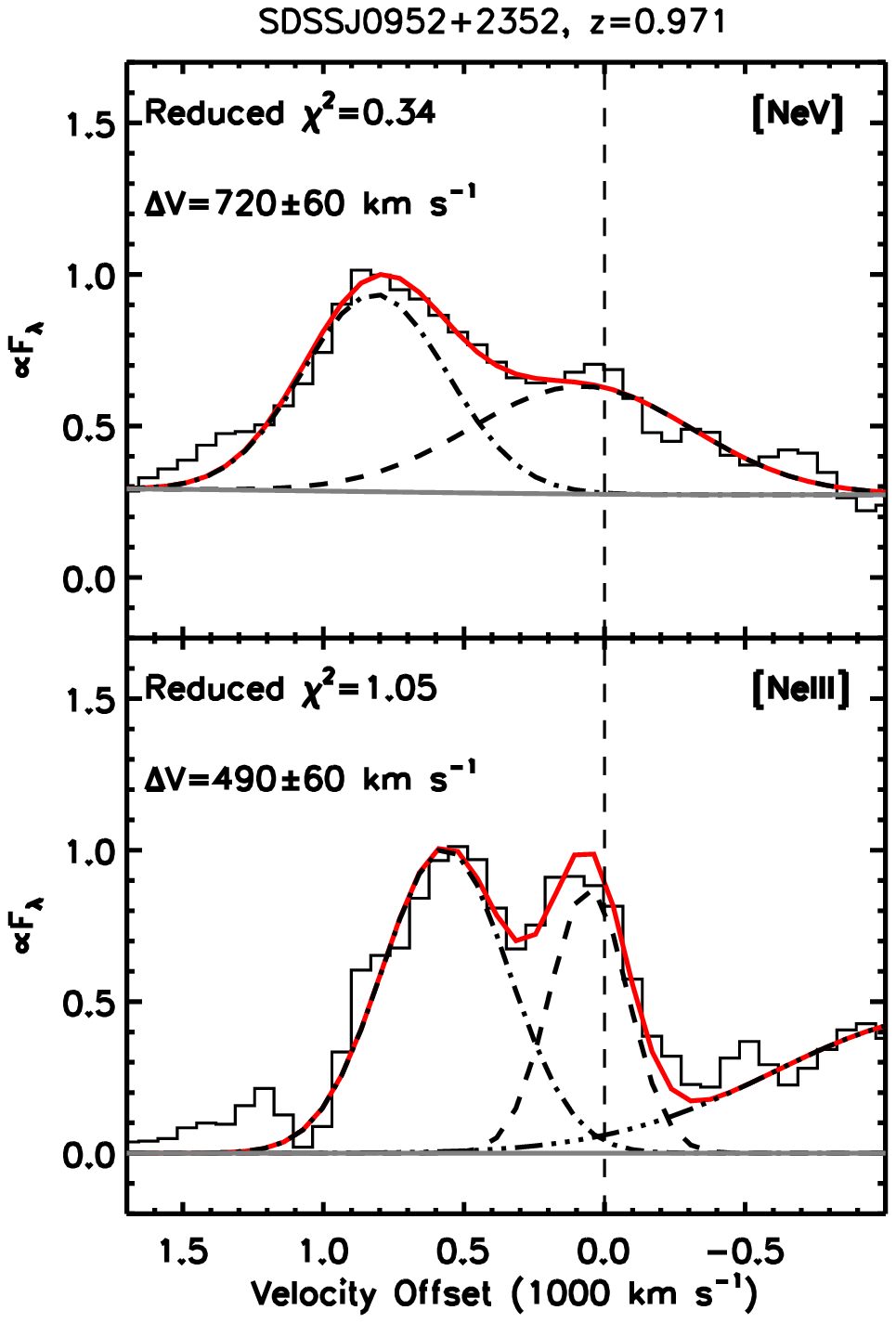} &
\hspace*{-0.1in} \includegraphics[width=1.8in, height=2.7in]{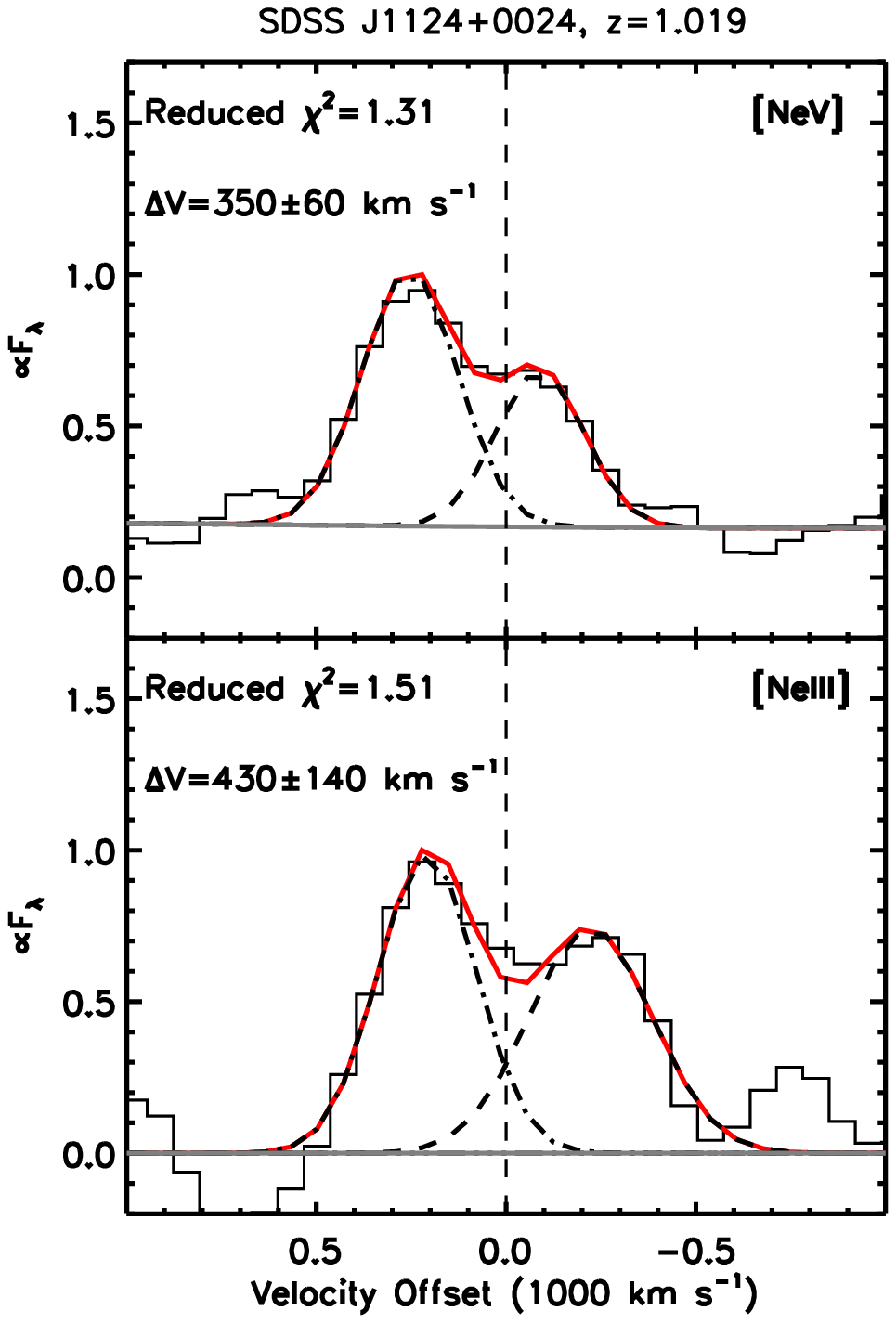}
\end{array} $
\caption{\footnotesize{Examples of two spectra from Figure \ref{examples} zoomed-in on \nev~(top) and \neiii~(bottom).  The spectra have been smoothed by convolving with a Gaussian of $\sigma=2$~\AA.  In each panel the peak flux of the double Gaussian model has been normalized to unity, and the spectra are plotted in velocity-space with zero-velocity at the quasar redshift.  The power-law continuum has been subtracted from the spectrum, and the model components are the same as described in Figure \ref{examples}.}}
\label{examples_close}
\end{figure}

\subsection{Modeling the Spectra and Selection of the Final Sample}
\label{modeling}
To generate the final sample to be used in our subsequent analyses, we modeled each spectrum yielded by the initial selection stage (Section \ref{sec:initial}) in order to determine if a two-Gaussian model significantly improves the line profile fit over a single-Gaussian model.  This modeling proceeded in two stages: 1) continuum modeling, and 2) emission line modeling.  

\subsubsection{Continuum Modeling}
\label{subsec:continuum}
The continuum was modeled by masking all detectable emission lines and simultaneously fitting a power-law function ($F_{\lambda}\sim \lambda^{-\alpha_{\lambda}}$) for the underlying AGN contiuum radiation plus a pseudo-continuum of broadened \feii~ emission lines from the empirical templates of \citet{Tsuzuki:2006} ($\lambda_{\rm{obs}} < 3500$~\AA) and \citet{veron_cetty:2004} ($\lambda_{\rm{obs}} > 3500$~\AA) which were developed from the spectrum of the narrow-line Seyfert 1 galaxy I Zw 1.  Due to the use of two separate \feii~templates, the powerlaw function was allowed to have a break at 3500~\AA, and the \feii~normalization allowed to vary independently below and above the break wavelength.  The \feii~pseudo-continuum is composed of many blended \feii~transitions which are generally believed to originate in or near the classical broad line region (BLR), and as such the redshift and broadening of the \feii~ emission should be comparable to that of the broad emission lines.  Therefore, in our modeling the \feii~emission was fixed at the redshift of the quasar being modeled (see Section \ref{redshifts} for a detailed discussion of the redshifts); we broadened the \feii~template by convolving with a Gaussian of $FWHM_{\rm{conv}}$ where $FWHM_{\rm{FeII}}^{2}=FWHM_{\rm{conv}}^{2}+FWHM_{\rm{I~Zw~1}}^{2}$ and $FWHM_{\rm{I~Zw~1}}=900$ km s$^{-1}$.  Use of this template necessarily limits the minimum $FWHM_{FeII}$ best-fit to 900 km s$^{-1}$, though this was not a problem since these sources were in the quasar catalog based on the presence of broad lines with $FWHM>1000$ km s$^{-1}$.  The lower rest-wavelength end of the continuum+\feii~fitting window was 2750 \AA~and the upper rest-wavelength end was 3950~\AA~(just redward of \neiii) if accessible, or otherwise the red end of the spectral coverage.  This fitting window allowed proper detection of the \feii~ emission since it has a relatively large equivalent width near \mgii.  We allowed $FWHM_{\rm{FeII}}$ to vary in steps of 100 km s$^{-1}$, and in general, the solutions are in the range $\sim$2000-9000 km s$^{-1}$, though in most cases the quality of the fits were not strongly dependent on the broadening.

\begin{figure}[t!]
\hspace*{-0.15in} \includegraphics[width=3.5in, height=3.15in]{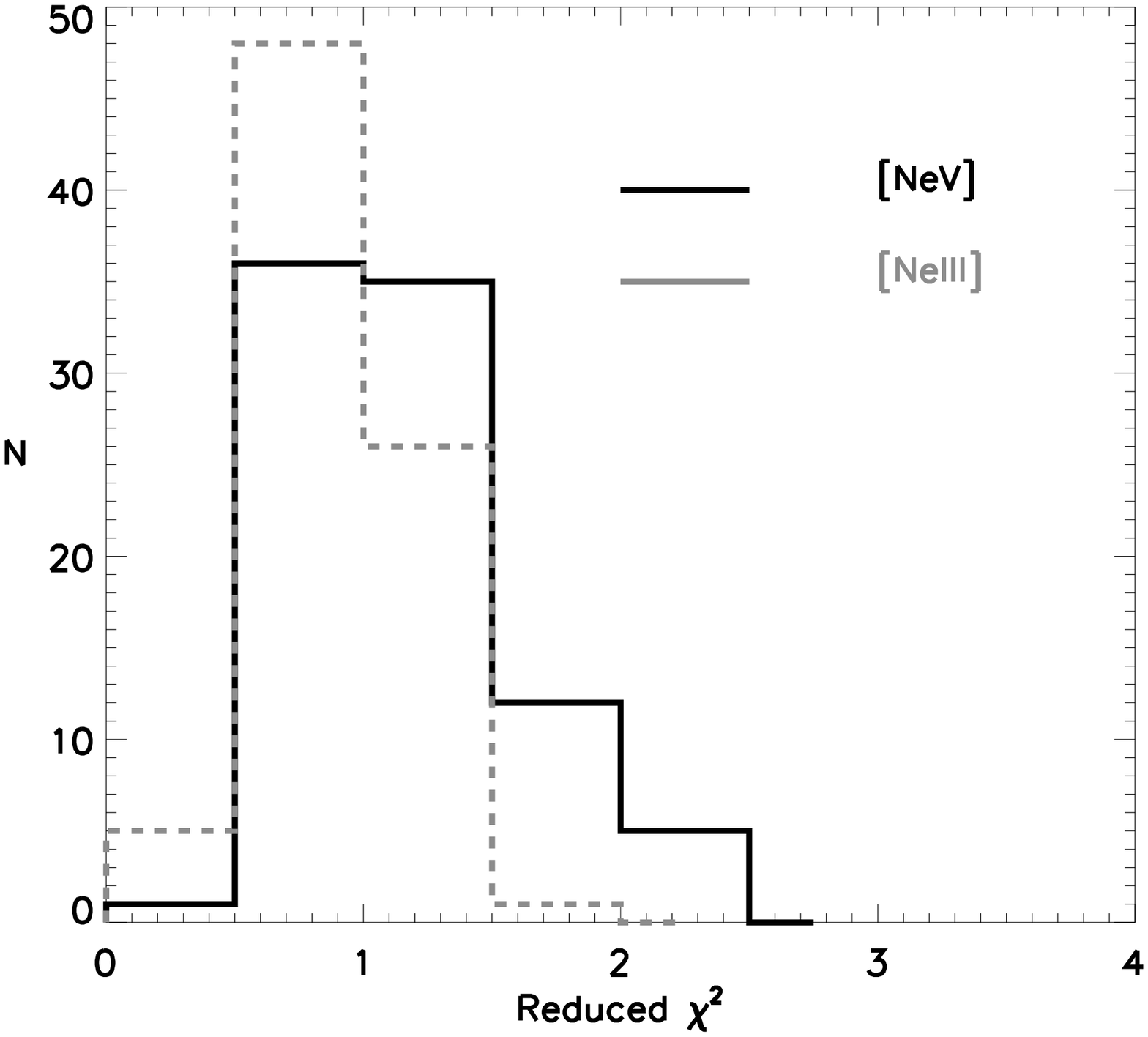} \\
\caption{\footnotesize{Distribution of reduced $\chi^{2}$ values for the best double-Gaussian fits for \nev~ (black, solid line) and \neiii~ (gray, dashed line).}}
\label{chi2}
\end{figure}

\subsubsection{Narrow Emission Line Modeling}
We fit each emission line (\nev~and \neiii) with both a single and a double Gaussian model, and required that sources in our final sample have fits to at least one of these two lines significantly improved by the double Gaussian model ($\Delta \chi^{2} \ge \Delta n_{DOF}$).  In seven individual line models, an additional broad component of line width $FWHM\approx700-1500$ km s$^{-1}$ was required for a satisfactory fit.  Three of the \nev~fits required such a component, where in two of those cases the broad component is consistent with the blue narrow peak (J074242.18+374402.0: $FWHM=1470$ km s$^{-1}$; and J105035.57+190544.2: $FWHM=1220$ km s$^{-1}$), and in the third case it is consistent with the red narrow peak (J150243.93+281739.9: $FWHM=1320$ km s$^{-1}$).  Four of the \neiii~fits required a broad component, where in three of those cases it is consistent with the blue narrow peak (J105035.57+190544.2: $FWHM=1520$ km s$^{-1}$; J105634.56+121023.5: $FWHM=820$ km s$^{-1}$; and J145659.27+503805.4: $FWHM=700$ km s$^{-1}$) and in one those cases it is located between the blue and red peaks (J085205.91+183922.2: $FWHM=1020$ km s$^{-1}$).  We note that in only one of these sources (J105035.57+190544.2) is a broad component seen in both \nev~and \neiii, and it is consistent with the blue peak.

Properly accounting for the contribution of flux from \feii~emission is often important in determining the profiles of the emission lines of interest in this wavelength range.  In particular, there is a local peak in the broadened \feii~emission just blueward of \nev, between 3390~\AA~and 3410~\AA, which has the potential to complicate the isolation of the blue \nev~component.  Therefore, careful attention was paid to the modeling in this region, and any sources for which the model was ambiguous were not admitted into the final sample.  However, there is no significant \feii~emission near \neiii, and only a small local peak redward of \oii~but with which it is not blended.

From our spectral modeling, there are 38 sources in our sample for which we have measured robust double-peaks in \emph{both} \nev~and \neiii, 42 sources for which we have \emph{only} robust \nev~measurements, and 51 sources for which we have \emph{only} robust \neiii~measurements.  Thus, there are a total of 131 double-peaked sources in our sample, making the fraction of double-peaked AGN detected in this manner $\sim$0.3\% of the parent sample.  See Figure \ref{examples} for examples of the spectra in our final sample showing the best-fit models over the rest-wavelength range containing \nev~and \neiii.  Additionally, Figure \ref{examples_close} compares the individual \nev~and \neiii~line profiles in velocity-space.  The best-fit parameters of the emission line modeling are listed in Table \ref{line_props1} (\nev) and Table \ref{line_props2} (\neiii).  All errors correspond to $1\sigma$ uncertainties and have been propagated throughout any further calculations.  The reduced $\chi^{2}$ values for each double Gaussian model are listed in Tables \ref{line_props1} and \ref{line_props2} and shown as a distribution in Figure \ref{chi2}.  The emission line modeling was performed with SPECFIT \citep{Kriss94}.  Our best-fitting \feii~FWHMs are listed in Table \ref{tab:general}, along with other relevant quasar properties (redshifts, \mgii~FWHMs, and Eddington ratios) which will be described in subsequent sections.

Throughout the rest of the paper, the velocity offsets of the individual blue and red components will refer to the offsets from the systemic velocity.  For \nev, these blue and red velocity offsets will be defined as  $\Delta V_{\rm{[NeV],blue}}$ and $\Delta V_{\rm{[NeV],red}}$, respectively.  For \neiii, the blue and red velocity offsets will be defined as $\Delta V_{\rm{[NeIII],blue}}$ and $\Delta V_{\rm{[NeIII],red}}$, respectively.  Blueshifts will correspond to positive velocities.  The velocity-splittings will then be the velocity difference between the lines, i.e. $\Delta V_{\rm{NeV}}\equiv \Delta V_{\rm{[NeV],blue}}-\Delta V_{\rm{[NeV],red}}$ and $\Delta V_{\rm{NeIII}}\equiv \Delta V_{\rm{[NeIII],blue}}-\Delta V_{\rm{[NeIII],red}}$.  The redshifts corresponding to the systemic velocity of each source are described in Section \ref{redshifts}. 

\subsection{Comparison Sample}
To generate a $z<0.8$ comparison sample to be used in our analyses, we examined the 89  \oiiis-selected Type 1 sources in  \citet{Smith:2010} and measured the line properties of double-peaked \nev~and \neiii~in cases for which two peaks are detectable in those lines.  We chose this sample because, as opposed to other double-peaked \oiii~samples, it includes Type 1 AGN which were selected from the SDSS quasar catalog.  Therefore, it provides a $z<0.8$ sample which can be used for direct comparison.  For the measurements on the \oiiis-selected sample, we selected those with robust \nevs/\neiiis~double-peaks in the same way as for our high-redshift sample, and likewise modeled the spectra as described above.  This resulted in 18 sources in the \oiiis-selected comparison sample for which we have measured robust double-peaks in \emph{both} \nev~and \neiii,  4 sources for which we have \emph{only} have robust \nev~measurements, and 20 sources for which we have \emph{only} have robust \neiii~measurements.  Thus, there are a total of 42 double-peaked sources in our comparison sample.

\section{Redshifts}
\label{redshifts}
Some sections of our analysis require a knowledge of the individual quasar redshifts.  For the sources in our sample, we have redshifts available from several different measurement techniques.  In this section we describe those redshift estimates which are useful for our scientific interests.  

\subsection{SDSS Redshifts}
Redshifts for spectroscopically-detected sources in the SDSS are determined by the `spectro1d' code described 

\begin{deluxetable*}{lccccccc}
\tabletypesize{\footnotesize}
\tablecolumns{8}
\tablecaption{Emission Line Properties for Double-peaked \nev}
\tablehead{
  \colhead{~} &
 \multicolumn{3}{c}{Blue} &
 \multicolumn{3}{c}{Red} \\ \hline
 \colhead{SDSS Name} \vline &
  \colhead{F$_{\lambda}$\tablenotemark{a}}&
 \colhead{$\Delta V$\tablenotemark{b}} &
 \colhead{FWHM\tablenotemark{c}} \vline &
  \colhead{F$_{\lambda}$\tablenotemark{a}} &
  \colhead{$\Delta V$\tablenotemark{b}} &
  \colhead{FWHM\tablenotemark{c}} \vline &
  \colhead{Reduced $\chi^{2}$ (dof)}
}
\startdata 
\hspace*{-0.in}J003159.87$+$063518.8 & $\phantom{00}  5.4\pm  4\phantom{00}$ & $  \phantom{0}  334\pm124  \phantom{0}$ & $ \phantom{0} 457\pm268 \phantom{0}$ & $\phantom{00}  6.9\pm  4\phantom{00}$ & $  \phantom{0}- 128\pm  6\phantom{000}$ & $ \ 465\pm203 \phantom{0}$ &   0.58 (10) \\
\hspace*{-0.in}J014933.86$+$143142.6 & $\phantom{00}  4.7\pm  2\phantom{00}$ & $  \phantom{0}  959\pm 55 \phantom{00}$ & $ \phantom{0} 260\pm 97\phantom{00}$ & $ \phantom{0} 18.7\pm  4\phantom{00}$ & $  \phantom{0-}  319\pm 57 \phantom{0}$ & $ \ 731\pm209 \phantom{0}$ &   1.25 (42) \\
\hspace*{-0.in}J015325.74$+$145233.4 & $ \phantom{0} 28.3\pm  8\phantom{00}$ & $  \phantom{0}  504\pm 49 \phantom{00}$ & $ \phantom{0} 403\pm119 \phantom{0}$ & $ \phantom{0} 33.9\pm 11 \phantom{0}$ & $  \phantom{0}- 176\pm 68 \phantom{00}$ & $ \ 585\pm249 \phantom{0}$ &   0.70 (32) \\
\hspace*{-0.in}J021648.36$-$092534.3 & $\phantom{00}  7.0\pm  2\phantom{00}$ & $  \phantom{0}  930\pm 35 \phantom{00}$ & $ \phantom{0} 245\pm 50\phantom{00}$ & $ \phantom{0} 34.9\pm  5\phantom{00}$ & $  \phantom{0-}  184\pm 41 \phantom{0}$ & $ \ 802\pm131 \phantom{0}$ &   0.65 (43) \\
\hspace*{-0.in}J021703.10$-$091031.1 & $ \phantom{0} 16.4\pm  4\phantom{00}$ & $  \phantom{0}  871\pm 35 \phantom{00}$ & $ \phantom{0} 364\pm 86\phantom{00}$ & $ \phantom{0} 28.0\pm  5\phantom{00}$ & $  \phantom{0-}  155\pm 51 \phantom{0}$ & $ \ 671\pm156 \phantom{0}$ &   0.82 (30) \\
\hspace*{-0.in}J034222.54$-$055727.9 & $ \phantom{0} 23.8\pm  6\phantom{00}$ & $  \phantom{0}  746\pm 73 \phantom{00}$ & $ \phantom{0} 641\pm206 \phantom{0}$ & $ \phantom{0} 24.4\pm  6\phantom{00}$ & $  \phantom{0}- 136\pm 83 \phantom{00}$ & $ \ 682\pm176 \phantom{0}$ &   0.90 (85) \\
\hspace*{-0.in}J035230.55$-$071102.3 & $  \phantom{}149.1\pm100  \phantom{}$ & $  \phantom{0}  690\pm270  \phantom{0}$ & $ \phantom{0} 933\pm267 \phantom{0}$ & $  \phantom{}244.1\pm108  \phantom{}$ & $ \phantom{00}-  15\pm110  \phantom{0}$ & $ \ 788\pm185 \phantom{0}$ &   0.72 (79) \\
\hspace*{-0.in}J073408.62$+$411901.1 & $ \phantom{0} 10.8\pm  4\phantom{00}$ & $  \phantom{0}  589\pm 81 \phantom{00}$ & $ \phantom{0} 364\pm214 \phantom{0}$ & $ \phantom{0} 20.1\pm  5\phantom{00}$ & $ \phantom{00}-  11\pm 68 \phantom{00}$ & $ \ 469\pm151 \phantom{0}$ &   1.95 (35) \\
\hspace*{-0.in}J074242.18$+$374402.0* & $\phantom{00}  9.9\pm  5\phantom{00}$ & $  \phantom{0}  198\pm 34 \phantom{00}$ & $ \phantom{0} 236\pm125 \phantom{0}$ & $ \phantom{0} 10.3\pm  4\phantom{00}$ & $  \phantom{0}- 116\pm 30 \phantom{00}$ & $ \ 209\pm 74\phantom{00}$ &   0.81 (33) \\
\hspace*{-0.in}J074641.70$+$352645.6 & $ \phantom{0} 38.9\pm 22 \phantom{0}$ & $   \phantom{} 1070\pm327  \phantom{0}$ & $  \phantom{}1306\pm476 \phantom{0}$ & $ \phantom{0} 22.8\pm 21 \phantom{0}$ & $  \phantom{0-}  124\pm254  \phantom{}$ & $ \ 961\pm373 \phantom{0}$ &   0.84 (58)
\enddata
\tablecomments{All errors correspond to $1\sigma$ uncertainties.  * denotes that a broad component was necessary.  This table is available in its entirety in a machine-readable form in the online journal. A portion is shown here for guidance regarding its form and content.}
\tablenotetext{a}{Flux densities, $F_{\lambda}$, are in units of $10^{-17}$ erg s$^{-1}$ cm$^{-2}$ \AA$^{-1}$.}
\tablenotetext{b}{$\Delta V$ is velocity offset from the quasar redshift in km s$^{-1}$.}
\tablenotetext{c}{FWHM are the observed line widths in units of km s$^{-1}$.}
\label{line_props1}
\end{deluxetable*}

\begin{deluxetable*}{lccccccc}
\tabletypesize{\footnotesize}
\tablecolumns{8}
\tablecaption{Emission Line Properties for Double-peaked \neiii}
\tablehead{
  \colhead{~} &
 \multicolumn{3}{c}{Blue} &
 \multicolumn{3}{c}{Red} \\ \hline
  \colhead{SDSS Name} \vline &
  \colhead{F$_{\lambda}$\tablenotemark{a}}&
 \colhead{$\Delta V$\tablenotemark{b}} &
  \colhead{FWHM\tablenotemark{c}} \vline &
  \colhead{F$_{\lambda}$\tablenotemark{a}} &
  \colhead{$\Delta V$\tablenotemark{b}} &
  \colhead{FWHM\tablenotemark{c}} \vline &
  \colhead{Reduced $\chi^{2}$ (dof)}
}
\startdata 
\hspace*{-0.in}J000531.41$+$001455.9 & $ \phantom{0} 28.0\pm  7\phantom{00}$ & $  \phantom{0}  146\pm 53 \phantom{00}$ & $ \phantom{0} 372\pm125 \phantom{0}$ & $ \phantom{0} 11.2\pm  5\phantom{00}$ & $  \phantom{0}- 179\pm 32 \phantom{00}$ & $ \ 125\pm 49\phantom{00}$ &   2.15 (30) \\
\hspace*{-0.in}J003159.87$+$063518.8 & $\phantom{00}  6.2\pm 11 \phantom{0}$ & $  \phantom{0}  446\pm139  \phantom{0}$ & $ \phantom{0} 223\pm278 \phantom{0}$ & $ \phantom{0} 18.1\pm 12 \phantom{0}$ & $ \phantom{00-}   61\pm232  \phantom{}$ & $ \ 560\pm274 \phantom{0}$ &   2.07 (31) \\
\hspace*{-0.in}J004305.92$-$004637.6 & $\phantom{00}  7.8\pm  6\phantom{00}$ & $  \phantom{0}  395\pm 74 \phantom{00}$ & $ \phantom{0} 236\pm276 \phantom{0}$ & $\phantom{00}  9.5\pm  6\phantom{00}$ & $  \phantom{0-}  111\pm 55 \phantom{0}$ & $ \ 209\pm151 \phantom{0}$ &   1.20 (32) \\
\hspace*{-0.in}J004312.70$+$005605.0 & $\phantom{00}  9.3\pm  2\phantom{00}$ & $  \phantom{0}  215\pm 31 \phantom{00}$ & $ \phantom{0} 215\pm 65\phantom{00}$ & $ \phantom{0} 16.6\pm  2\phantom{00}$ & $  \phantom{0}- 140\pm 13 \phantom{00}$ & $ \ 212\pm 24\phantom{00}$ &   1.32 (33) \\
\hspace*{-0.in}J014822.62$+$132142.7 & $ \phantom{0} 37.8\pm 15 \phantom{0}$ & $  \phantom{0}  883\pm 78 \phantom{00}$ & $  \phantom{}1048\pm438 \phantom{0}$ & $ \phantom{0} 27.6\pm  6\phantom{00}$ & $  \phantom{0}- 376\pm102  \phantom{0}$ & $ \ 779\pm149 \phantom{0}$ &   0.83 (47) \\
\hspace*{-0.in}J015734.24$+$003405.5 & $ \phantom{0} 17.1\pm  5\phantom{00}$ & $  \phantom{0}  806\pm 50 \phantom{00}$ & $ \phantom{0} 260\pm 93\phantom{00}$ & $ \phantom{0} 70.6\pm 12 \phantom{0}$ & $ \phantom{00}-  69\pm 42 \phantom{00}$ & $ \ 672\pm131 \phantom{0}$ &   1.05 (37) \\
\hspace*{-0.in}J021648.36$-$092534.3 & $\phantom{00}  8.9\pm  2\phantom{00}$ & $  \phantom{0}  538\pm 27 \phantom{00}$ & $ \phantom{0} 185\pm 50\phantom{00}$ & $ \phantom{0} 35.8\pm  5\phantom{00}$ & $\phantom{000}-   4\pm 36 \phantom{00}$ & $ \ 546\pm 95\phantom{00}$ &   1.06 (24) \\
\hspace*{-0.in}J023234.33$-$091053.0 & $ \phantom{0} 51.4\pm 17 \phantom{0}$ & $   \phantom{} 1462\pm157  \phantom{0}$ & $  \phantom{}1565\pm506 \phantom{0}$ & $ \phantom{0} 34.4\pm 10 \phantom{0}$ & $ \phantom{00-}   31\pm151  \phantom{}$ & $  1121\pm238 \phantom{0}$ &   0.73 (66) \\
\hspace*{-0.in}J034222.54$-$055727.9 & $ \phantom{0} 28.8\pm  9\phantom{00}$ & $  \phantom{0}  480\pm  0\phantom{000}$ & $ \phantom{0} 925\pm302 \phantom{0}$ & $ \phantom{0} 58.4\pm  7\phantom{00}$ & $  \phantom{0}- 323\pm 37 \phantom{00}$ & $ \ 656\pm 81\phantom{00}$ &   1.09 (78) \\
\hspace*{-0.in}J035230.55$-$071102.3 & $  \phantom{}120.0\pm 42 \phantom{0}$ & $  \phantom{0}  382\pm 68 \phantom{00}$ & $ \phantom{0} 433\pm135 \phantom{0}$ & $  \phantom{}201.2\pm 40 \phantom{0}$ & $ \phantom{00}-  70\pm 33 \phantom{00}$ & $ \ 379\pm 57\phantom{00}$ &   0.95 (32)
\enddata
\tablecomments{All labels are the same as in Table \ref{line_props1}.  This table is available in its entirety in a machine-readable form in the online journal. A portion is shown here for guidance regarding its form and content.}
\label{line_props2}
\end{deluxetable*}

\begin{deluxetable*}{lccccc}
\tabletypesize{\footnotesize}
\tablecolumns{6}
\tablecaption{General Quasar Properties for Our Sample}
\tablehead{
 \colhead{SDSS Name} &
  \colhead{z$_{\rm{SDSS}}$} &
  \colhead{z$_{\rm{MgII}}$} &
 \colhead{FWHM$_{\rm{FeII}}$\tablenotemark{a}} &
 \colhead{FWHM$_{\rm{MGII}}$\tablenotemark{a,b}} &
 \colhead{f$_{\rm{Edd}}$\tablenotemark{c}}
}
\startdata
J000531.41$+$001455.9 & $0.9918\pm 0.0009$ & $0.9931\pm 0.0018$ & 9000 & $ 5424.76\pm  1192.59$ & 0.109 \\
J003159.87$+$063518.8 & $1.0921\pm 0.0010$ & $1.0935\pm 0.0016$ & 1000 & $ 2475.19\pm  1745.19$ & 0.085 \\
J004305.92$-$004637.6 & $0.8482\pm 0.0015$ & $0.8488\pm 0.0015$ & 8400 & $ 2828.74\pm   247.77\phantom{0}$ & 0.094 \\
J004312.70$+$005605.0 & $0.9036\pm 0.0010$ & $0.9047\pm 0.0014$ & 8500 & $ 6223.22\pm   283.39\phantom{0}$ & 0.105 \\
J014822.62$+$132142.7 & $0.8767\pm 0.0018$ & $0.8759\pm 0.0019$ & 8000 & $ 6063.22\pm   134.05\phantom{0}$ & 0.146 \\
J014933.86$+$143142.6 & $0.9024\pm 0.0013$ & $0.9027\pm 0.0013$ & 4500 & $ 2969.71\pm   244.58\phantom{0}$ & 0.172 \\
J015325.74$+$145233.4 & $1.1755\pm 0.0020$ & $1.1762\pm 0.0021$ & 5100 & $ 5219.56\pm   553.29\phantom{0}$ & 0.202 \\
J015734.24$+$003405.5 & $1.0834\pm 0.0013$ & $1.0825\pm 0.0020$ & 3300 & $ 6846.37\pm  1636.32$ & 0.109 \\
J021648.36$-$092534.3 & $0.8834\pm 0.0013$ & $0.8834\pm 0.0014$ & 2700 & $ 4288.12\pm   632.18\phantom{0}$ & 0.119 \\
J021703.10$-$091031.1 & $0.8752\pm 0.0013$ & $0.8757\pm 0.0015$ & 2700 & $ 2598.31\pm   432.35\phantom{0}$ & 0.289
\enddata
\tablecomments{This table is available in its entirety in a machine-readable form in the online journal. A portion is shown here for guidance regarding its form and content.}
\tablenotetext{a}{FWHMs are in units of km s$^{-1}$.}
\tablenotetext{b}{Values for $FWHM_{MgII}$ are from the SDSS DR7 Catalog of Quasar Properties \citep{Shen:2011a}.}
\tablenotetext{c}{Eddington ratio calculations are described in Section \ref{ledd}.}
\label{tab:general}
\end{deluxetable*} 

\noindent in \citet{Stoughton:2002}.  In short, two separate redshifts are determined, an emission line redshift ($z_{EL}$) and a cross-correlation redshift ($z_{XC}$), and the final value adopted by the code is the redshift solution with the highest confidence level (though each individual value is stored and available through the archive).  Additionally, a small fraction of the quasar redshifts were re-determined after visual inspection as described in \citet{Schneider:2010}.  $z_{EL}$ is measured by the identification of common galaxy and quasar emission lines with known rest-wavelength values, and therefore for our sample the highest confidence $z_{EL}$ values are almost exclusively determined by sets of the strong quasar emission lines \civ, \ciii, \mgii, \oii, H$\delta$, H$\gamma$, and H$\beta$.  Likewise, the template which provided the highest confidence $z_{XC}$ values for our sample is the composite quasar template from \citet{Berk:2001}.  For 107 of our sources, $z_{XC}$ and $z_{EL}$ are consistent with each other within their errors.  For the remaining 24 redshifts, 19 are from $z_{XC}$, 4 are from $z_{EL}$, and 1 was determined by hand.  We note that for those sources with disagreeing values of $z_{XC}$ and $z_{EL}$, the poorer of the two values is clearly incorrect and not usable.  Throughout the rest of this paper, we refer to the final redshifts produced by the `spectro1d' code as $z_{SDSS}$, and these values are listed in Table \ref{tab:general}.

\begin{figure}[t!]
\hspace*{-0.15in} \includegraphics[width=3.5in, height=3.15in]{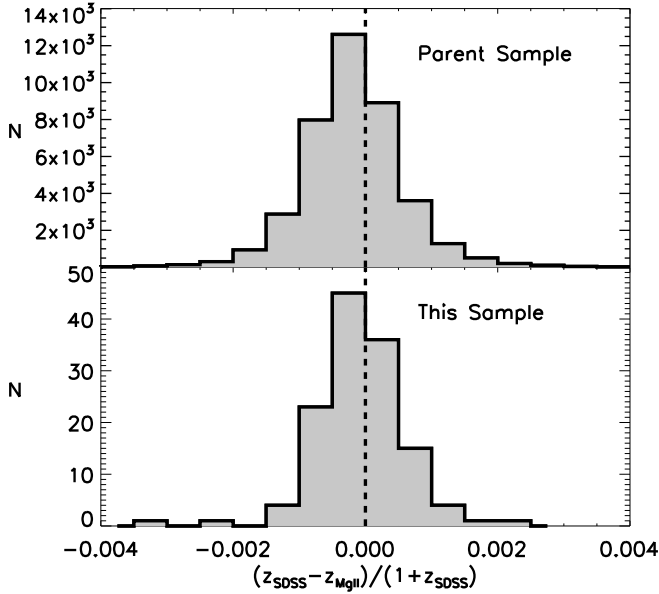}
\caption{\footnotesize{Distribution of $(z_{SDSS}-z_{MgII})/(1+z_{SDSS})$ for the parent sample (top) and our sample (bottom).  The mean and sigma are shown for each distribution.  Note that the distributions are similar for both samples.}}
\label{fig:z_offsets}
\end{figure}

\subsection{Redshifts from \mgii}
\label{subsec:z_MGII}
In principle, individual emission lines can provide independent redshift estimates corresponding to the physical regions where those lines originated, e.g. the broad emission lines provide redshifts for the BLR.  For our purposes, it is useful to know the redshift of the central SMBH (under the assumption of a single, active SMBH in the host galaxy) which can potentially be traced by the BLR if the gas is virialized.  Fortuitously, given the redshift range of our sample, for all of our sources we have spectral access to the broad \mgii~emission line, which has been shown to be virialized in the potential of the central SMBH \citep{McLure:Jarvis:2002}.  Therefore, we have used the \mgii~emission line models from the catalog of SDSS DR7 quasar properties \citep{Shen:2011a}  to obtain estimates of the BLR redshifts ($z_{MgII}$).  

\begin{figure}[t!]
\hspace*{-0.2in} \includegraphics[width=3.5in, height=3.15in]{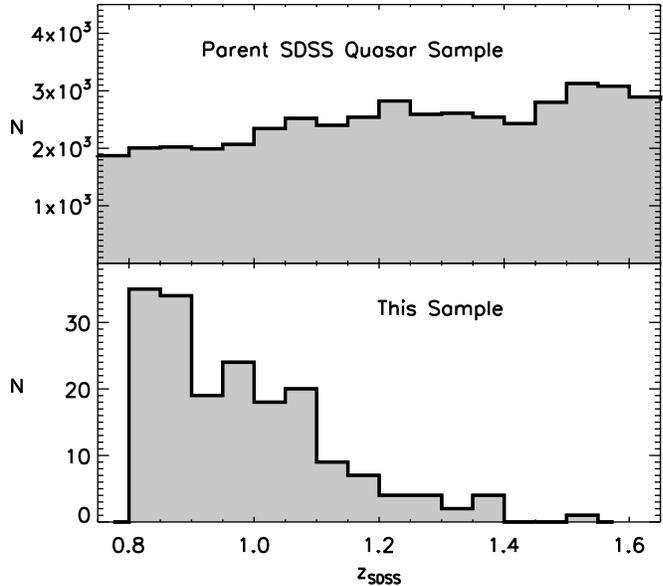}
\caption{\footnotesize{Top: Distribution of the parent SDSS quasar sample \citep{Schneider:2010} in the relevant redshift range.  Bottom: Distribution of redshifts for our final double-peaked \nevs/\neiiis~sample developed as described in Section \ref{analysis}.  The cut-off below $z=0.80$ in our sample is entirely the result of our initial cut in redshift space (see Section \ref{sample}), and the highest redshift for sources found through our selection process is $z=1.53$.  The redshifts are described in Section \ref{redshifts}.}
}
\label{fig:redshifts}
\end{figure}

However, we would like to note several potential technical difficulties involved in measuring the \mgii~line centroid which limit its application to our work.  These difficulties include blending with \feii~emission, UV absorption features, and the often ambiguous presence of a narrow emission line component (which is actually the doublet \ion{Mg}{2}$\lambda\lambda2797,2802$).  We also note that the \mgii~line centroids may be complicated if the low-ionization broad emission lines have the characteristic double-peaked profile of disk emitters \citep{EH03,strateva03}.  Furthermore, since we are considering the scenario of two AGN within the same host galaxy, we must consider the possibility that the broad \mgii~line is a blending of two BLRs from two Type 1 AGN.  Though the components of a dual AGN would not be close enough for the line-splitting to exceed the FWHM (several thousand km s$^{-1}$) and produce explicit double-peaked profiles \citep{Shen:2010a}, two broad \mgii~components separated by several hundred km s$^{-1}$ will still result in a relatively broadened, and possibly asymmetric, \mgii~profile, thereby complicating the centroid measurement.  We have visually inspected the \mgii~line profile for each source in our sample in order to characterize their structure.  There are several sources which show evidence for asymmetric structure, though it is generally difficult to discern if any of the profiles contain two broad components as would be the case if there are dual BLRs.  Though many of those redshifts have rather large errors, a subset of them have robust line centroids with percent errors less than $1\%$ (73 sources) and percent errors less than $0.1\%$ (52 sources).  The \mgii~redshifts are listed in Table \ref{tab:general}.

\subsection{Comparison of $z_{SDSS}$ and $z_{MgII}$}
Figure \ref{fig:z_offsets} shows the distribution of $(z_{SDSS}-z_{MgII})/(1+z_{SDSS})$ where the \mgii~redshifts appear to be systematically larger than the cross correlation redshifts in both our sample and the parent sample.  This same effect is clearly apparent in the SDSS redshift analysis of \citet{Hewett:2010}, which suggests that we are seeing the same systematic trend in our analysis.  While \citet{Hewett:2010} provide redshifts which correct for this systematic offset in a statistical sense, these corrections may not be appropriate for some individual quasars.  Therefore, we do not use these improved redshifts since our sample only contains 131 sources.  

Based upon our comparison between $z_{SDSS}$ and $z_{MgII}$, for the rest of our analysis we have chosen to use the $z_{SDSS}$ values and their associated errors.  These SDSS redshifts will serve as the systemic velocities for our sources.  All quoted line-splittings between the two line peaks ($\Delta V_{\rm{[NeV]}}$ and $\Delta V_{\rm{[NeIII]}}$) and velocity offsets for the blue ($\Delta V_{\rm{[NeV],blue}}$ and $\Delta V_{\rm{[NeIII],blue}}$) and red ($\Delta V_{\rm{[NeV],red}}$ and $\Delta V_{\rm{[NeIII],red}}$) components are relative to these SDSS redshifts.  The range of redshifts in our sample is $z=0.80$ (low-limit cutoff discussed in Section \ref{sample}) to $z=1.53$ (highest redshift source in our sample).  In Section \ref{sec:dynamics}, for which our analysis is heavily dependent upon the choice of redshift, we will also use the robust $z_{MgII}$ measurements mentioned in Section \ref{subsec:z_MGII} for comparison. 

\section{General Properties of the Sample}
\label{sec:offsets}
In this section we describe several of the general properties of our sample.  We compare the distributions of these properties (redshifts, S/N, and velocity offsets) to the parent sample or our low-redshift comparison sample in order to show our selection biases.  Additionally, we discuss the results of our completeness estimates with respect to several of these and other properties.  

\subsection{Distributions}
The SDSS redshifts we have adopted for our sample (Section \ref{redshifts}) are shown and compared to the parent sample in Figure \ref{fig:redshifts}.  The redshift distribution shows a peak near the low redshift cutoff of the sample, with a gradual decline out to higher redshifts, in contrast to the increasing population of the parent SDSS quasar sample.  This strong dependence on redshift may reflect a dependence on equivalent width or a luminosity bias, which we discuss further below.

\begin{figure}[t!]
\hspace*{-0.2in} \includegraphics[width=3.5in, height=3.15in]{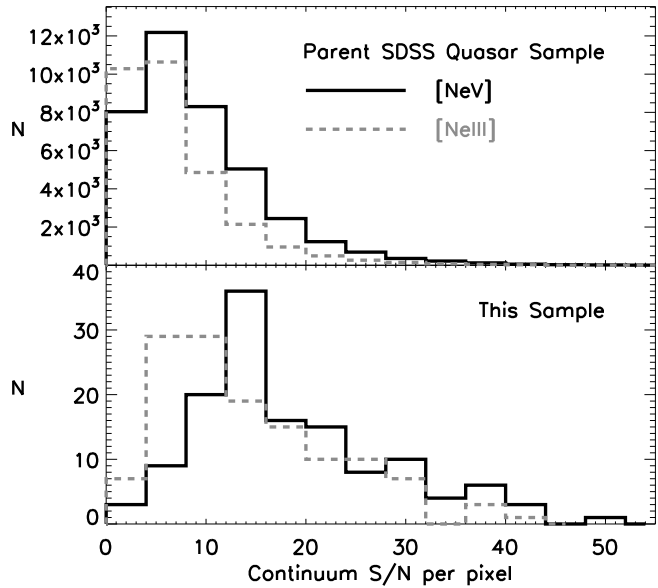} \\
\caption{\footnotesize{Distribution of continuum S/N per pixel near the \nev~(black, solid line) and \neiii~(gray, dashed line) emission lines for sources in our sample.  For each distribution, the S/N values were calculated over a set of featureless wavelength regions near the respective emission lines.}
}
\label{SN}
\end{figure}

Figure \ref{SN} shows the distribution of the continuum S/N per pixel, where the mean values are 18.5 and 14.5 in the wavelength regions adjacent to \nev~and \neiii, respectively.  Compared to the parent sample, our sample shows a deficit at the low S/N regime, and a stronger tail out to higher S/N values.  These distributions show that our selection is biased toward spectra of good quality which reflect the fact that detection of double-peaked structure in \nev~and \neiii~is sensitive to the S/N and generally difficult at S/N$<$5.  Furthermore, the larger mean value of the continuum S/N near \nev~is reflective of the fact that \nev~is generally a weaker line than \neiii~in AGN spectra \citep{Osterbrock:2006} and therefore a stronger signal is required for the detection of double-peaked structure.  The dependence of our selection completeness on S/N will be addressed below.

Since this study is motivated by the identification of high-redshift analogs of AGN with double-peaked \oiii~lines, we compare the distribution of line properties of our sample to those of the \oiiis-selected comparison sample.  The SDSS spectral resolution, S/N, and the intensity of the emission lines determine whether or not we are able to reliably detect double-peaks at a given separation in velocity-space.  \nev~and \neiii~are fainter than \oiii~in AGN spectra by a factor of $\sim4$ \citep{Ferland:Osterbrock:1986}, which means that selection based on \nevs/\neiiis~will be biased toward larger line-splittings relative to \oiii.  For example, double-peaked samples selected through \oiii~have sources with line splittings down to $\sim$200 km s$^{-1}$, and the mean of the \oiii~line-splittings in the Type 1 sample of \citet{Smith:2010} is $\sim$420 km s$^{-1}$.  On the other hand, the mean of the \nev~and \neiii~line-splittings of our sample is $\sim$700 and $\sim$700 km s$^{-1}$, respectively.  

\begin{figure*}[t!]
\hspace*{0.1in} \includegraphics[width=7.in, height=3.5in]{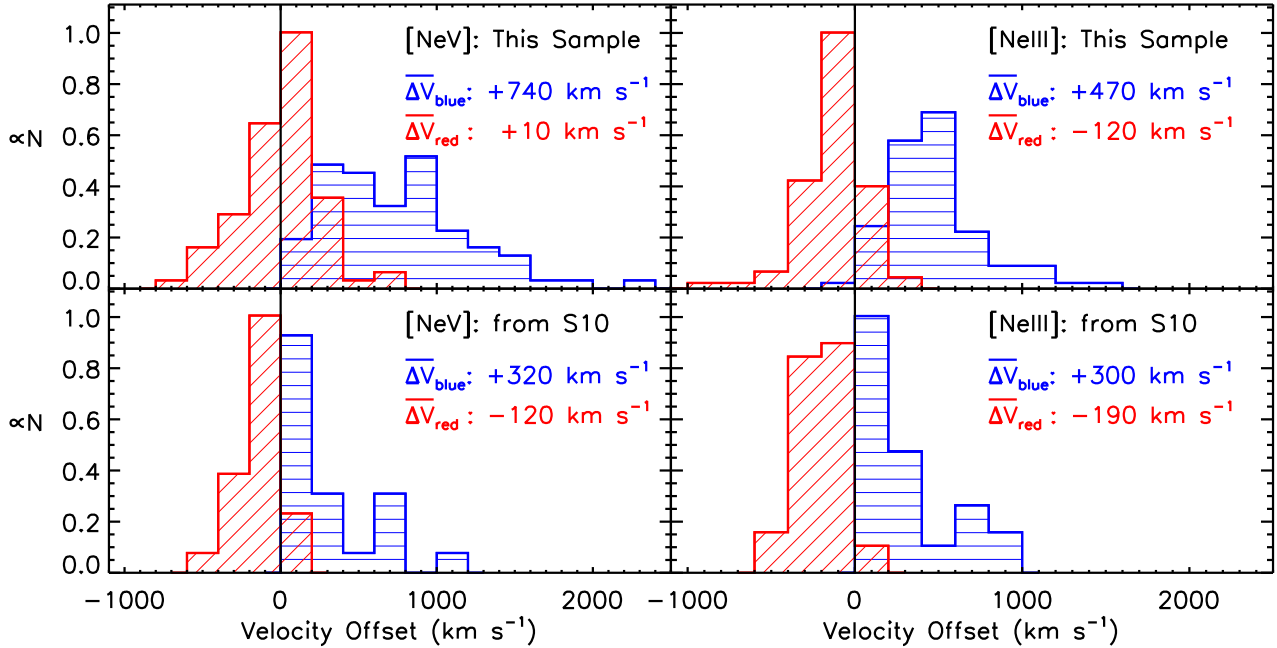} \\
\caption{\footnotesize{Distribution of velocity offsets for double-peaked AGN with positive velocities indicating blueshifts.  `Blue' component offsets ($\Delta V_{\rm{[NeV],blue}}$ and $\Delta V_{\rm{[NeIII],blue}}$) are shown as horizontally-hatched blue lines, and `red' component offsets ($\Delta V_{\rm{[NeV],red}}$ and $\Delta V_{\rm{[NeIII],red}}$) are shown as diagonally-hatched red lines.  The peak value in each panel has been normalized to unity.  Top: Sources selected through \nev~and \neiii~emission line profiles.  Bottom: Distribution of velocity-offsets for the \oiiis-selected Type 1 AGN from the \citet{Smith:2010} (S10) sample.  All distributions are shown separately for \nev~(left) and \neiii~(right).  The average velocity-offsets are shown for each distribution.  The black solid vertical line in each panel represents zero velocity offset.}
}
\label{offsets}
\end{figure*}

Figure \ref{offsets} shows the distribution of velocity offsets from the systemic redshift for \nev~and \neiii~in both our sample and the \oiiis-selected sample.  While the measurements on the comparison sample were performed in the same way as for our sample (Section \ref{analysis}), these sources were selected from a sample of previously determined double-peaked AGN.  Therefore, we were able to more confidently identify two peaks in \nev~and \neiii~at relatively smaller \dv s for the \oiiis-selected sample, as shown in Figure \ref{offsets}.  A Kolmogorov-Smirnov (KS) test indicates that the blue components of our sample come from a different distribution than those of the \oiii-selected sample ($D_{\rm{[NeV]}}=0.56$, $D_{\rm{[NeIII]}}=0.43$, and probabilities that they come from the same distribution $P_{\rm{[NeV]}}=3\times 10^{-5}$, $P_{\rm{[NeIII]}}=7\times 10^{-5}$).  A KS test also reveals a (less significant) difference in the red components ($D_{\rm{[NeV]}}=0.43$, $D_{\rm{[NeIII]}}=0.33$, and probabilities of $P_{\rm{[NeV]}}=2.3\times 10^{-3}$, $P_{\rm{[NeIII]}}=5.6\times 10^{-3}$).  This indicates that we are finding similar double-peaked narrow line sources to those selected through \oiii~but that we are preferentially missing a portion of those with relatively small \dv 's.

\subsection{Selection Completeness}
\label{completeness}
To estimate what fraction of sources with double-peaked \nev~and/or \neiii~we are missing due to the potential selection biases discussed above, we have estimated our selection completeness through simulations.  Specifically, we generated artificial spectra designed to mimic SDSS spectra in the \nevs/\neiiis~wavelength ranges (a similar test was performed by \citealt{Liu2010a} for their \oiiis-selected sample).  Using the SDSS spectral dispersion, we modeled the continuum as a power-law and the emission lines as Gaussians with randomly assigned values for the parameters of equivalent width (EW), FWHM, continuum S/N per pixel, and line offsets from the systemic velocity.  The ranges of allowed values for those parameters were made uniform and at least slightly larger than the distributions of our final sample in order to ensure that we sampled the relevant parameter space for the analysis.  While the EW distribution of our sample only extended to $\sim$35~\AA~for the blue and red components, we allowed our simulations to have EWs of up to 50~\AA~for the individual lines (100~\AA~total EW).  The simulated FWHMs ranged from 100 to 1800 km s$^{-1}$, $\sim$200 km s$^{-1}$ larger than the largest FWHM of our sample.  The simulated continuum S/N values ranged from 2 to 33, just larger than the range of our sample (see Figure \ref{SN}).  Finally, the simulated line offsets ranged from 0 to 2000 km s$^{-1}$, $\sim$500 km s$^{-1}$ larger than the largest line offsets of our sample.  We generated a total of 10,000 spectra, with 2,000 of them having double-peaked emission lines, and the other 8,000 having single-peaked emission lines.  Out of these simulated spectra, randomly distributed, we selected a final sample of double-peaked emission lines in the same manner as described in Section \ref{analysis}. 

We see a positive trend between completeness and EW, and find that we are highly incomplete over the EW range of our sample.  For example, we estimate a completeness of just $\sim$1.2$\%$ for a total EW of 10~\AA~and $\sim$4.5$\%$ for a total EW of 30~\AA.  We see a negative trend between completeness and FWHM which can be interpreted as the result of peak blending at large line widths.  We see a moderate positive trend between completeness and S/N, with a completeness of $\sim$42$\%$ at S/N=5 and $\sim$50$\%$ at S/N=30.  Finally, we see a strong dependence on velocity-splitting, with a completeness of $\sim$55\% at $\Delta V=900$ km s$^{-1}$ and $\sim$34\% at $\Delta V=500$ km s$^{-1}$.    In contrast, \citet{Liu2010a} find a completeness of $\sim$75\% at $\Delta V=900$ km s$^{-1}$ and $\sim$50\% at $\Delta V=500$ km s$^{-1}$ for their \oiii~ sample.  
 
 Over comparable parameter ranges, we are less complete than \citet{Liu2010a} for all of the parameters ($\sim$50-80\% of their completeness), with the exception of FWHM, for which we have an equivalent or slightly greater completeness at the large FWHM end of their distribution ($\sim$900 km s$^{-1}$).  This is likely due to the fact that we allowed for a larger range of line offsets resulting in our ability to identify broader double peaks.  The smaller completeness in our selection compared to those of \citet{Liu2010a} is likely due to the relatively smaller equivalent widths of \nevs/\neiiis~allowed in our simulations which is reflective of the weaker intensities compared to \oiii.  Of the parameters investigated, we find that our completeness is most strongly dependent on and most drastically different from \citet{Liu2010a} in the velocity-offsets (Figure \ref{offsets}).  This result is again consistent with the notion that we are preferentially missing a portion of those with relatively small \dv s.

Since false double-peaks are generated by noise, false positives are most likely to occur in low S/N spectra in which random noise peaks are difficult to discern from the emission line signal.  However, our conservative selection criteria effectively required that the S/N be sufficiently large, as can be seen in Figure \ref{SN}, with $S/N_{\rm{[NeV]}}<5$ for only 4 sources (3\%) and $S/N_{\rm{[NeIII]}}<5$ for only 11 sources (8\%).  This is reflected in the fact that we did not select any false positives in our completeness analysis.

\begin{figure}[t!] $
\begin{array}{c c}
\hspace*{-0.2in} \includegraphics[width=3.5in,height=5in]{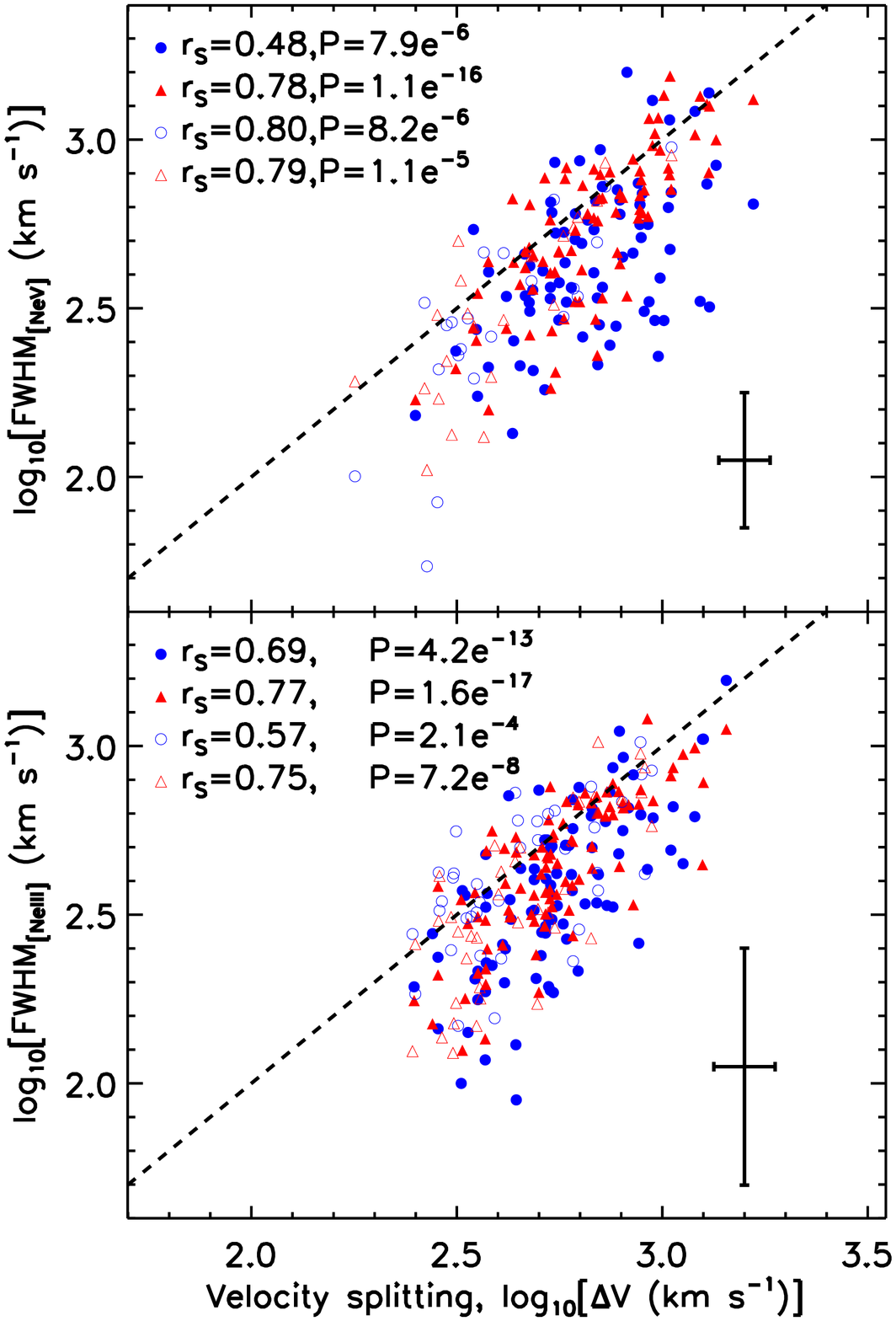}
\end{array} $
\caption{\footnotesize{Plot of line width (FWHM) vs. line-splitting for \nev~(top panel) and \neiii~(bottom panel).  In each panel, the `blue' and `red' systems are labeled with blue circles and red triangles, respectively.  Filled data points correspond to our \nevs/\neiiis-selected sample, while unfilled points correspond to the \oiiis-selected sample.  The Spearman rank coefficients ($r_{S}$) and probabilities of a null hypothesis ($P$) are listed for the `blue' and `red' components.  The dashed line is the one-to-one relation.  The average X and Y errors are indicated in the lower right corner of each panel.  Our completeness tests indicate that we are increasingly \emph{incomplete} toward large-FWHM/small-\dv~ (i.e. the upper left corners), but that we are increasingly \emph{complete} toward small-FWHM/large-\dv~ (i.e. the lower right corners).}}
\label{relv_width}
\end{figure} 

\section{Correlations among Sample Properties}
\label{kinematics}
In the following section, we examine several trends related to the velocity offsets of the individual blue and red components from the systemic redshift and line-splittings between the blue and red components.  We are interested in uncovering the prevailing mechanism(s) producing the double-peaked emission lines in our sample.  In particular, we investigate the following four relationships: line-splitting vs line width, line-splitting vs quasar Eddington ratio, \nev~ velocity offsets vs \neiii~ velocity offsets, and blue/red velocity-offset ratio vs blue/red luminosity ratio.  In the following analyses we combine the \nevs/\neiiis~ measurements of our sample with those of the \oiiis-selected sample in order to increase both the sample size and the dynamic range of velocity offsets and line-splittings.

\subsection{A Correlation Between Line-Splittings and Line Widths}
\label{sec:relv_width}
In AGN with emission lines offset from the systemic velocity, there is often an observed positive correlation between the line peak offset (blueshifts being positive offsets) and line width.  For example, in samples of so-called `blue outliers' (AGN with \oiii~lines blueshifted by $>100$ km s$^{-1}$), there is an observed positive correlation between the blueshift and line width of \oiii~\citep{Komossa:2008b}.  A similar trend was also seen by \citet{Liu2010a} for their sample of Type 2 double-peaked \oiii~sources, with the offsets being the line-splitting between the two peaks.  In a similar fashion, we have plotted the line splittings against the line widths (FWHM) for the double-peaked \nev~and \neiii~lines in both our sample and in the \oiiis-selected sample (Figure \ref{relv_width}).  For both \nev~and \neiii, statistically significant positive correlations between the line-splitting and FWHM are evident.  The correlations are strong and the line-splittings are generally larger than the line widths, and both of these results were also found by \citet{Liu2010a} with respect to \oiii~emission lines.  

\begin{figure*}[t!] $
\begin{array}{cc}
\hspace*{-0.1in}\includegraphics[width=7in,height=5.6in]{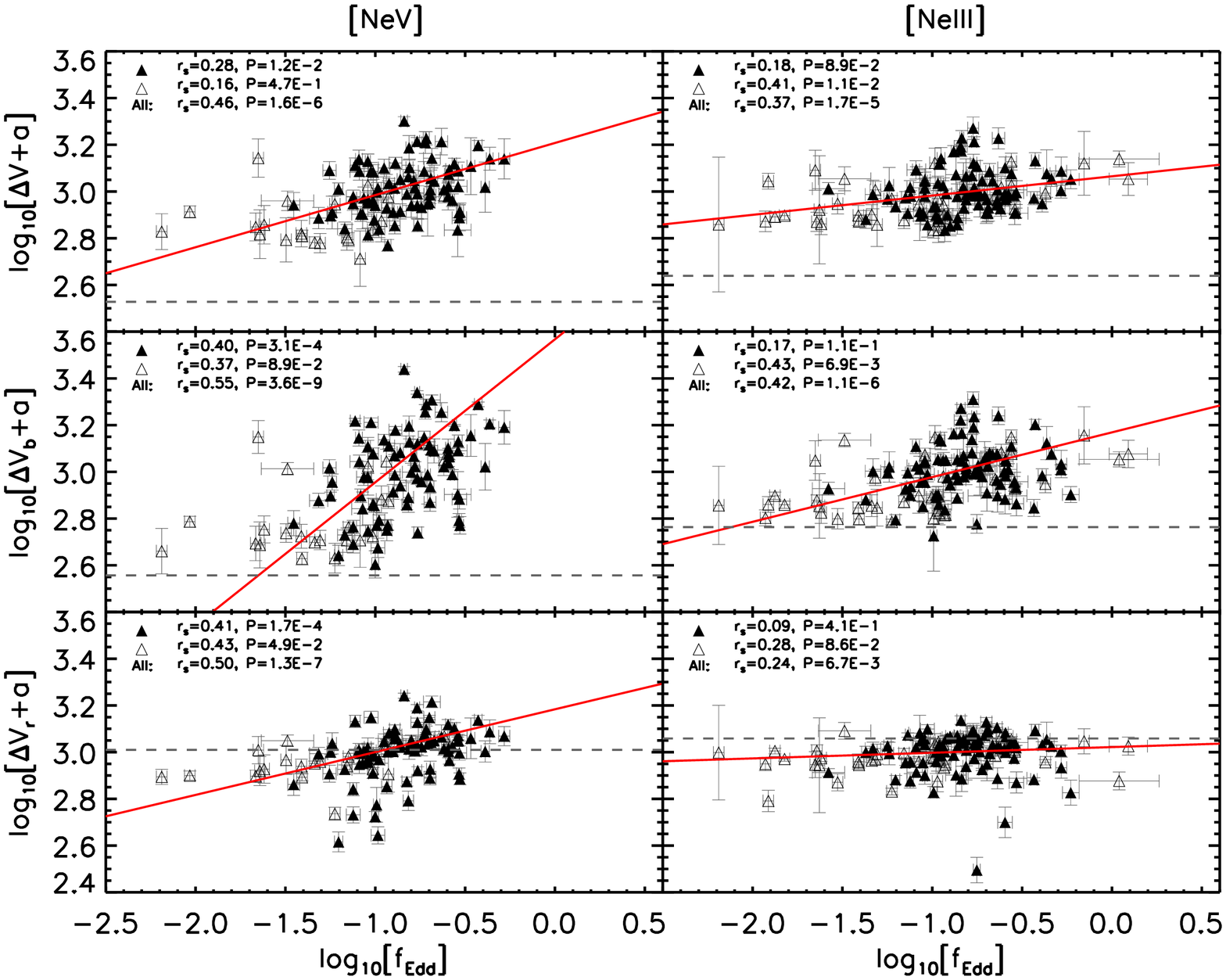}
\end{array} $
\caption{\footnotesize{Plot of Eddington ratio versus velocity-splitting (top), blue component offsets (middle) and red component offsets (bottom) for \nev~ (left) and \neiii~ (right).  See Section \ref{ledd} for details on calculations of $f_{\rm{Edd}}$.  To be consistent among each subsample while accommodating the negative velocity-offsets of red components, the individual velocities for each distribution have been shifted by a constant, $a$, such that the mean of that particular distribution is equivalent to 1000 km s$^{-1}$, i.e. $a=1000-(\overline{\Delta V})$.  In each panel the horizontal gray, dashed line represents $\Delta V=0$ km s$^{-1}$ for that particular distribution.  Sources at $0.8<z<1.6$ (\nevs/\neiiis-selected) are plotted as filled triangles, and sources at $z<0.8$ (\oiiis-selected) are plotted as unfilled triangles.  The Spearman rank coefficients and probabilities of the null hypothesis are given for each of the sub-samples and for the combined samples.  The red solid lines in each distribution are the best linear fits weighted by the X and Y errors (obtained with the `FITEXY' routine) for the combined sample.}}
\label{fig:ledd}
\end{figure*}

The possible physical interpretations of these trends often includes an ionization stratification of the NLR in which the higher ionization lines originate closer to the ionizing radiation source \citep{Zamanov:2002,Komossa:2008b}.  In such a scenario, if there is a radially decelerating outflowing component of the NLR, then the higher ionization lines are accelerated to higher velocities.  The emission lines produced in the inner portions of the NLR will also be more broadened as their motion would be dominated by the bulge gravitational potential \citep{Nelson:Whittle:1996}.  Since the ionizing potentials of \nev~and \neiii~($I.P.=41.07$ and $97.16$ eV, respectively) are larger than that of \oiii~($I.P.=35.15$ eV), we might expect the correlation between line-splitting and FWHM to be stronger among these emission lines.  This is also expected based on the higher critical electron densities for collisional de-excitation ($n_{\rm{crit}}$) of \nev~and \neiii~($n_{\rm{crit}}=1.3\times 10^{7}$ cm$^{-3}$ and $n_{crit}=9.5\times 10^{6}$ cm$^{-3}$, respectively) compared to \oiii~($n_{\rm{crit}}=6.8\times 10^{5}$ cm$^{-3}$) since electron densities will increase toward the nuclear region.  Indeed, we find that the strengths of these correlations are generally comparable to or stronger than those for \oiii~in the sample of \citet{Liu2010a} and in the narrow line Seyfert 1 (NLS1) sample from \citet{Komossa:2008b}.  These trends in our sample might be more reflective of the even stronger trends among the subset of `blue-outliers' from \citet{Komossa:2008b}.

\begin{figure*}[t!] $
\begin{array}{c c}
\hspace*{-0.2in} \includegraphics[width=3.5in, height=2.4in]{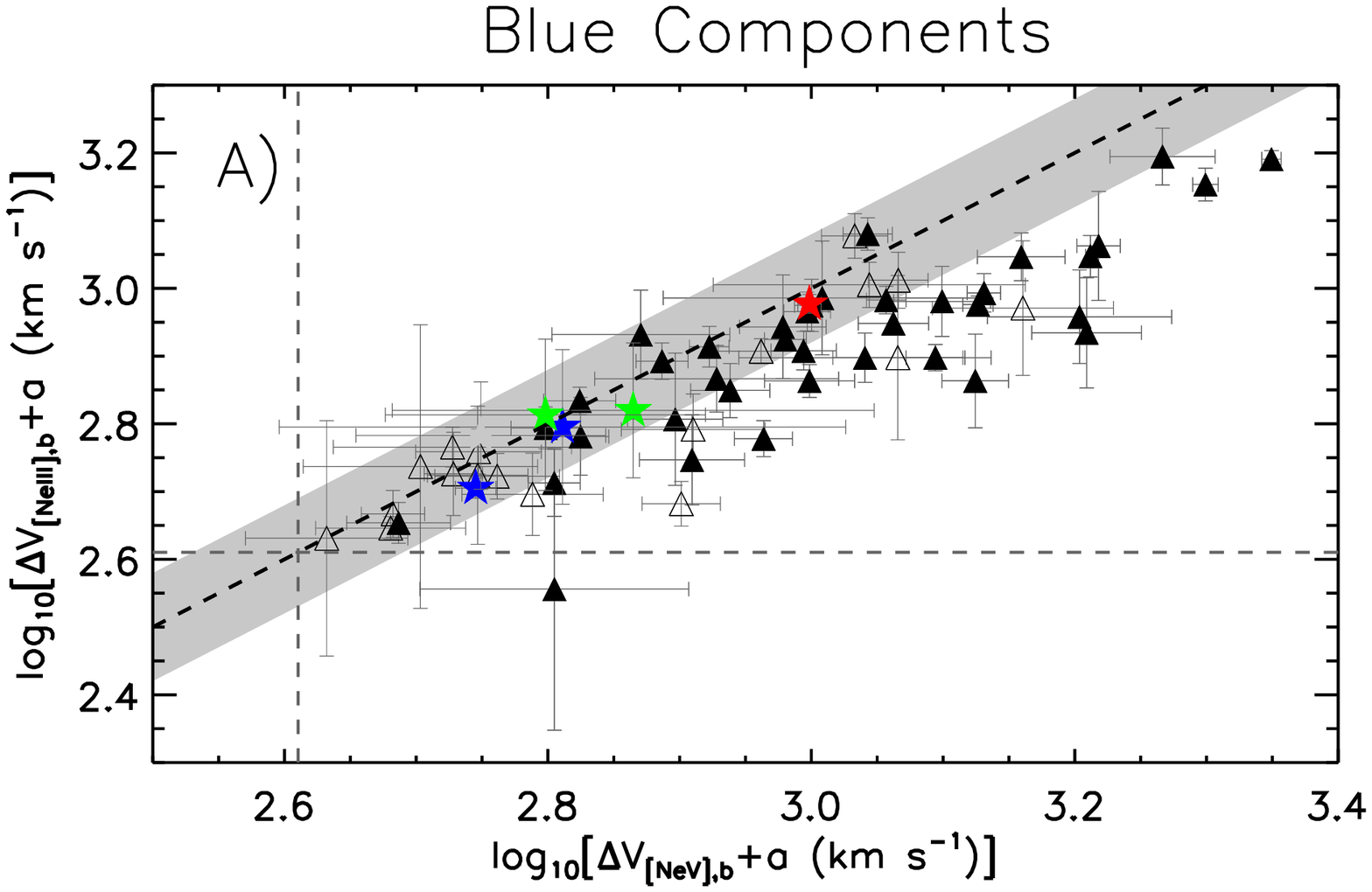} &
\includegraphics[width=3.5in, height=2.4in]{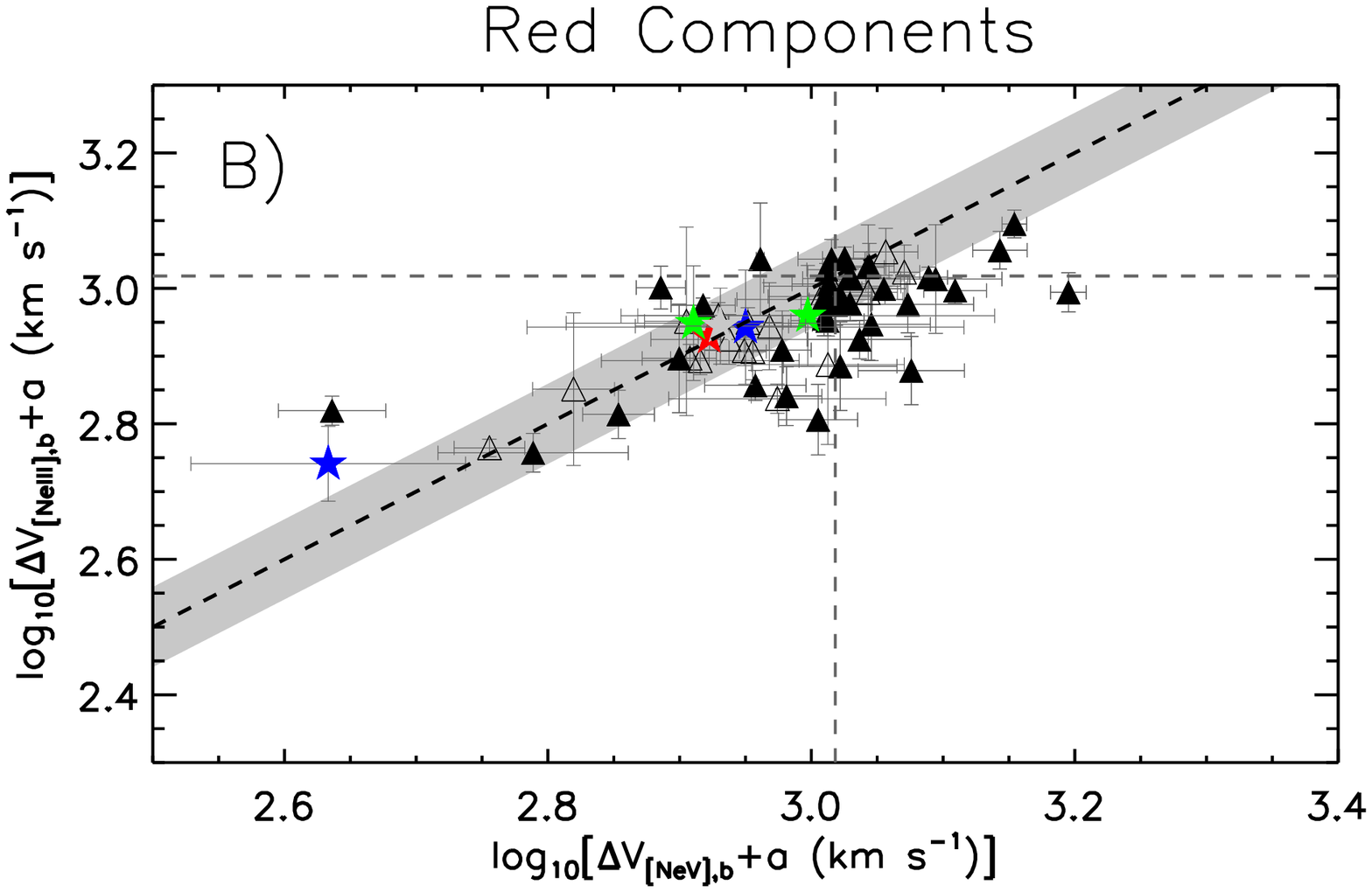} \\
\hspace*{-0.2in} \includegraphics[width=3.5in, height=2.4in]{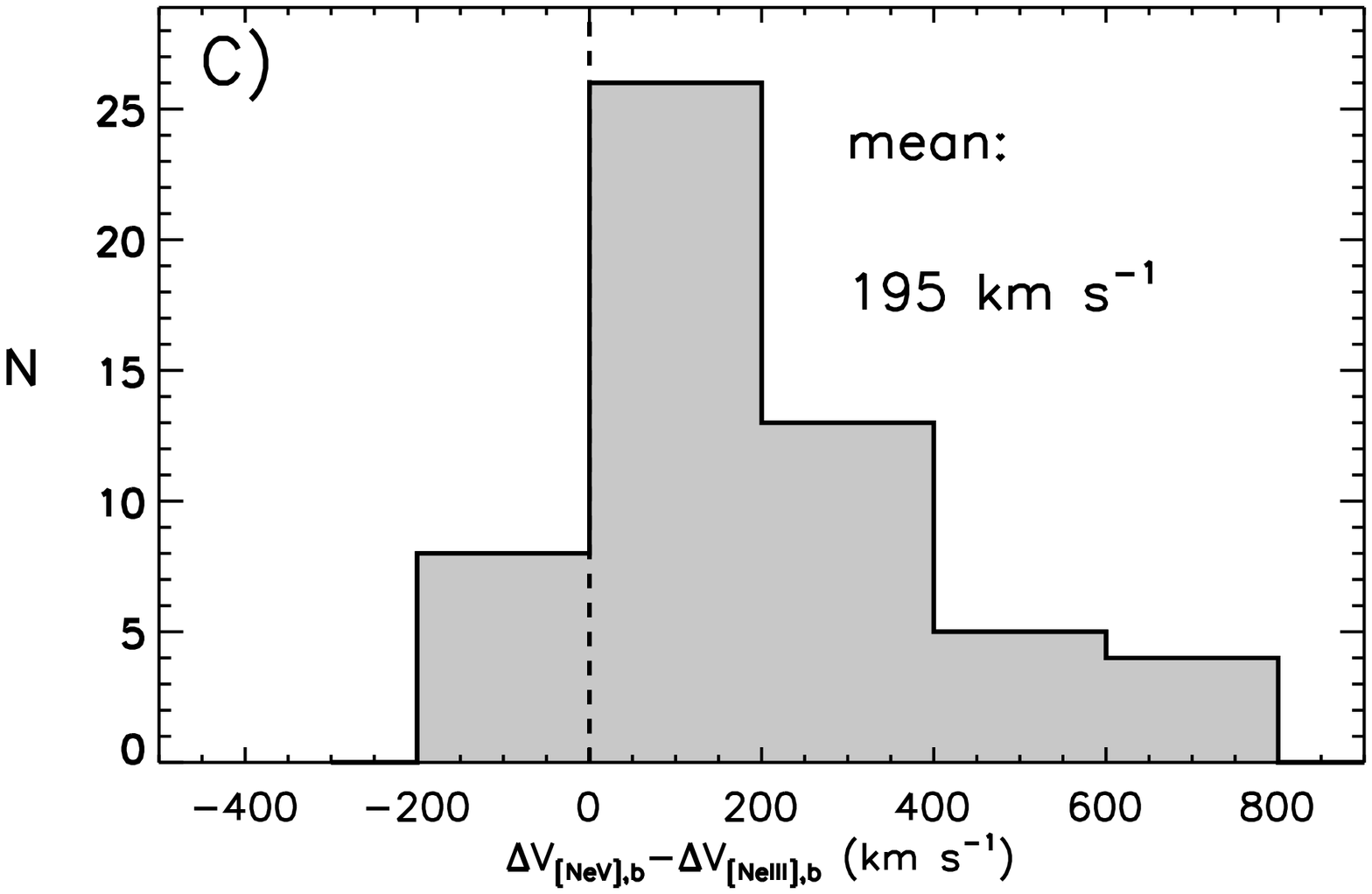} &
\includegraphics[width=3.5in, height=2.4in]{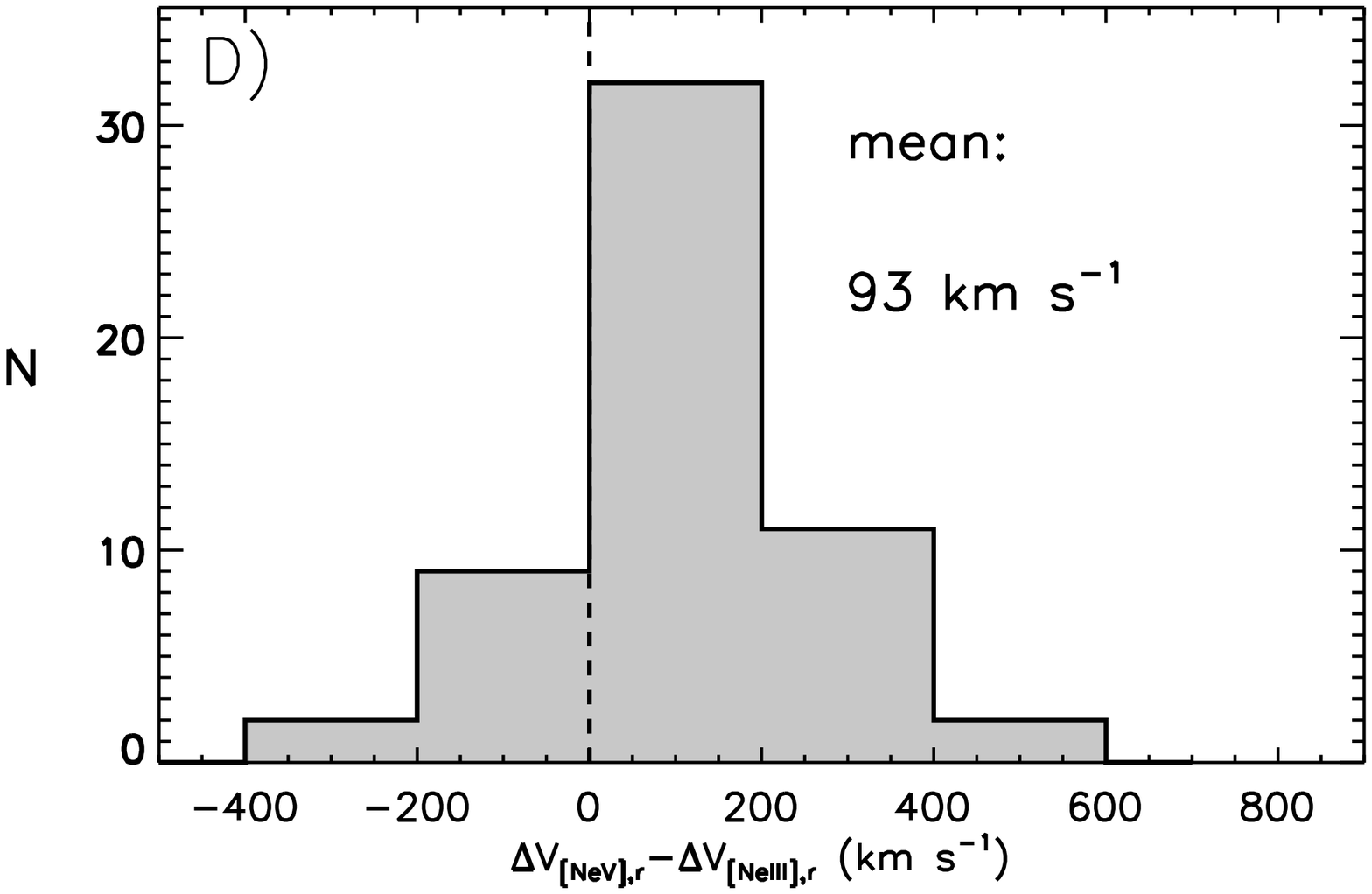}
\end{array} $
\caption{\footnotesize{Top: Plots of $\Delta V_{\rm{blue, [NeV]}}$ versus $\Delta V_{\rm{blue, [NeIII]}}$ (A) and $\Delta V_{\rm{red, [NeV]}}$ versus $\Delta V_{\rm{red, [NeIII]}}$ (B).  The dashed line is the one-to-one relation.  Each velocity has been shifted by the constant $a$ where $a=1000-(\overline{\Delta V_{[NeV]}})$.  In each panel the horizontal gray, dashed line represents $\Delta V_{\rm{[NeV]}}=0$ km s$^{-1}$ and the gray, shaded region represents the average of our sample.  \nevs/\neiiis-selected sources are plotted as filled circles while sources selected via \oiii~ are plotted as open circles.  Additional sources are plotted as stars: extended NLRs (SDSS J110851.03+065901.4 and SDSS J135646.10+102609.0) and an unresolved NLR (SDSS J105052.46+083934.7) from \citet{Fu:2012} are indicated by solid green and purple stars, respectively, [OIII]-selected dual AGN SDSS J171544.05+600835.7 \citep{Comerford:2011} and SDSS J150243.1+111557 \citep{Fu:2011} are indicated by blue stars, and a \nevs/\neiiis-selected dual AGN candidate from \citet{Barrows:2012} is indicated by a red star (CXOXBJ142607.6+353351).  Bottom: Histograms of $\Delta V_{\rm{[NeV],blue}}-\Delta V_{\rm{[NeIII],blue}}$ (C) and $\Delta V_{\rm{[NeV],red}}-\Delta V_{\rm{[NeIII],red}}$ (D).  The dashed line in the bottom panels represents $\Delta V_{\rm{[NeV],blue}}=\Delta V_{\rm{[NeIII],blue}}$ (C) and $\Delta V_{\rm{[NeV],red}}=\Delta V_{\rm{[NeIII],red}}$ (D).  The mean values are shown for each distribution.}}
\label{fig:strat}
\end{figure*}

\subsection{A Correlation Between Velocity-Splitting and Eddington Ratio}
\label{ledd}
Motivated by the outflow interpretation often used for samples of double-peaked emission line AGN, we investigate the relationship between the velocity-splittings and the quasar Eddington ratios ($f_{\rm{Edd}}=L_{\rm{bol}}/L_{\rm{Edd}}$).  We calculated $L_{\rm{Edd}}$ using the standard derivation of the Eddington luminosity, $L_{\rm{Edd}}=4\pi cGM_{\rm{BH}}\mu_{e}/\sigma_{T}$, where $G$ is the gravitational constant, $\mu_{e}$ is the mass per unit electron, and $\sigma_{T}$ is the Thomson scattering cross-section \citep{Krolik:1999}.  $M_{\rm{BH}}$ was estimated for each Type 1 AGN using the SMBH mass-scaling relationships from broad emission line widths and monochromatic luminosities: $FWHM_{\rm{H\beta}}$ and $L${\scriptsize5100~\AA} ($z<0.8$) or $FWHM_{\rm{MgII}}$ and $L${\scriptsize3000~\AA} ($z>0.8$) from McClure and Dunlop (2004).  The line widths are mostly from the SDSS DR7 catalog of quasar properties \citet{Shen:2011a}, though we visually inspected and re-fit several of them for which we were able to improve on the models.  For each AGN with an estimate of $M_{\rm{BH}}$, we estimated $L_{\rm{bol}}$ from the monochromatic luminosities  $L${\scriptsize5100~\AA} ($z<0.8$) or $L${\scriptsize3000~\AA} ($z>0.8$) and the bolometric corrections of Richards et al. (2006).  

Figure \ref{fig:ledd} shows $f_{\rm{Edd}}$ plotted against three quantities: velocity-splittings, blue velocity-offsets and red velocity-offsets.  For both \nev~and \neiii, a Spearman rank test reveals statistically significant correlations between $f_{\rm{Edd}}$ and each of these quantities.  When comparing \nev~and \neiii, it is apparent that the correlations are stronger for \nev~than for \neiii~in all correlations tested.  Additionally, for both \nev~ and \neiii~the correlations are strongest between $f_{\rm{Edd}}$ and the blue velocity-offsets.  We note that the low velocity-offsets are dominated by the \oiiis-selected sources, which simply reflects the selection bias discussed in Section \ref{sec:offsets}.  If the velocity-splittings represent a NLR outflow in a significant number of these sources, then this result suggests that the radiation generated by the quasar accretion rate may play a crucial role in driving the line-splittings.  Additionally, it appears that the blue component is more strongly dominated by outflowing material than the red component, which has implications for obscuration.

\begin{figure*}[t!] $
\begin{array}{cc}
\hspace*{-0.1in}\includegraphics[width=7in,height=4.2in]{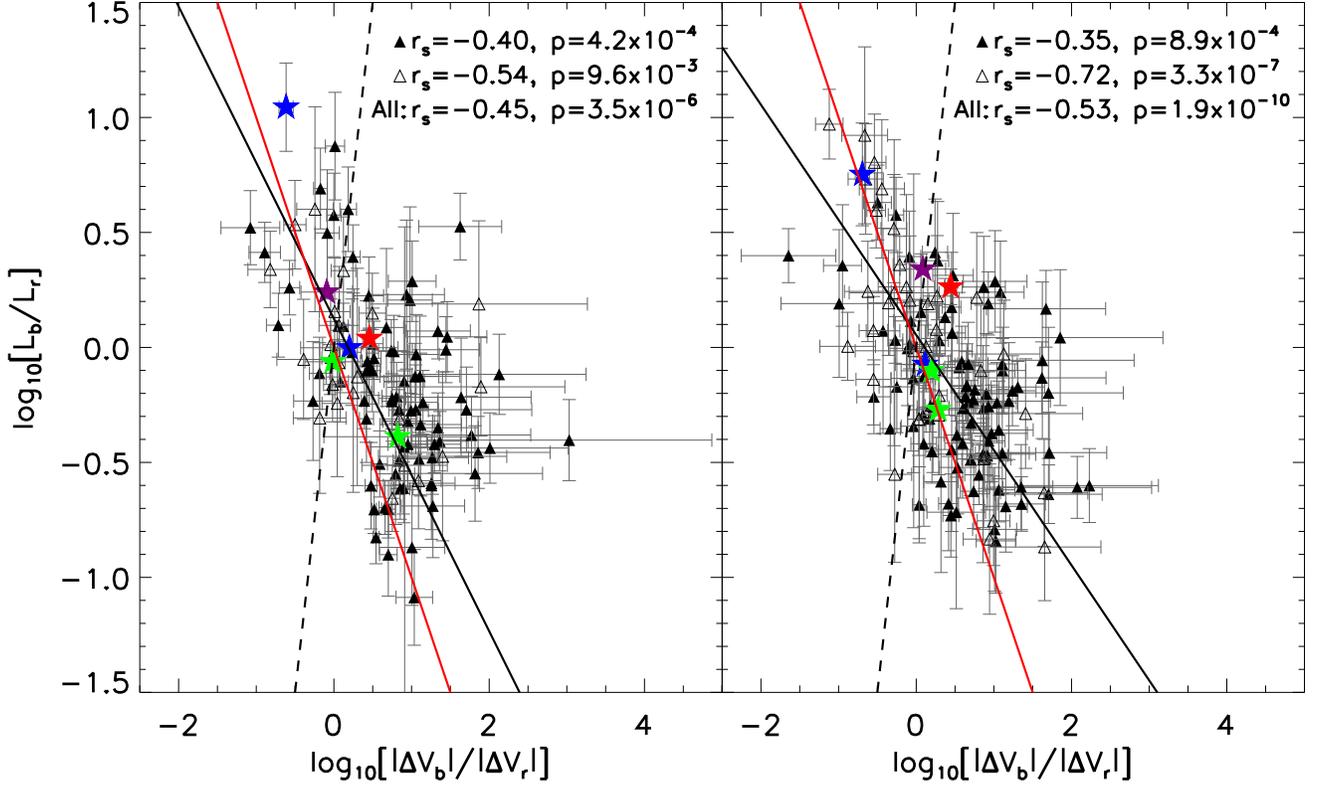}
\end{array} $
\caption{\footnotesize{Plot of the velocity-offset ratio ($|\Delta V_{b}|/|\Delta V_{r}|$) versus the luminosity ratio ($L_{b}/L_{r}$) shown separately for \nev~(left) and \neiii~ (right).  The solid black line is linear best fit weighted by the X and Y errors found using the `FITEXY' routine.  The solid red line is the theoretical relation assuming equivalent Eddington ratios for the blue and red components ($\epsilon_{1,2}=1$), and the dashed black line is the theoretical relation for a bi-polar outflow, both of which are discussed in Section \ref{sec:dynamics}.  The colored stars represent the same extended/unresolved NLRs, confirmed dual AGN and candidate dual AGN as in Figure \ref{fig:strat} and have the same symbols.  In each panel the Spearman rank correlations coefficients and probabilities of no correlation are shown for our sample, this comparison sample, and the combined (`All') sample.}}
\label{fig:flux_shift}
\end{figure*}

\subsection{Ionization Stratification}
\label{variation}
The difference in I.P. between \nev~and \neiii~is $\sim$56 eV and can be used to provide insight on the gas dynamics in different regions of the NLR.  Since the line offsets have been shown to correlate with the quasar Eddington ratios (Section \ref{ledd}), suggesting an outflowing component, we would like to search for evidence of an ionization stratification.  Figures \ref{fig:strat}A and \ref{fig:strat}B show $\Delta V_{\rm{[NeV], blue}}$ versus $\Delta V_{\rm{[NeIII], blue}}$ and $\Delta V_{\rm{[NeV],red}}$ versus $\Delta V_{\rm{[NeIII],red}}$ for all sources in our sample and the \oiiis-selected sample for which there are detectable double-peaks in both lines.  

Though the true nature of the gas kinematics in NLRs is likely to be complex, under the simplest pictures of NLR ionization stratifications it is generally not expected that lines of lower ionization potential will show blue velocity-offsets larger than those of greater ionization potentials.  Indeed, from Figure \ref{fig:strat}A it can be seen that all of the points are consistent with having equivalent blue velocity-offsets in the two lines, or otherwise with a larger blue offset in \nev.  This is also apparent in Figure \ref{fig:strat}C, with a mean $\Delta V_{\rm{[NeV],blue}}-\Delta V_{\rm{[NeIII],blue}}$ value of 195 km s$^{-1}$.  Accounting for the $1 \sigma$ uncertainties, from our sample 15 sources (40\%) are consistent with equivalent blue velocity-offsets for both \nev~and \neiii, and 23 sources (60\%) have significantly larger \nev~blue velocity offsets, consistent with the presence of an ionization stratification.  Comparable fractions are seen in the \oiiis-selected sample.  The sources in the small velocity-offset portion of the plot are dominated by the \oiiis-selected sample, which again reflects the selection bias discussed in Section \ref{sec:offsets}.  In contrast, the difference in the red offsets has a narrower distribution and a mean value more consistent with $\Delta V_{\rm{[NeV],red}}=\Delta V_{\rm{[NeIII],red}}$ (Figures \ref{fig:strat}B and \ref{fig:strat}D).  This suggests that the red systems are generally less stratified and have a different origin.  

Also plotted are several additional double-peaked sources for which follow-up observations have been obtained, including extended and unresolved NLRs from \citet{Fu:2012}, dual AGN from \citet{Fu:2012} and \citet{Comerford:2011}, and a candidate dual AGN from \citep{Barrows:2012}.  Note that the extended/unresolved NLRs and the dual AGN are \oiiis-selected, and the candidate dual AGN was selected based on \nev~and \neiii.  With the exception of the dual AGN SDSS J150243.1+111557, these additional sources are all consistent with no apparent ionization stratification in either the blue or red system by the $1\sigma$ criteria we have used above.

\subsection{Dynamical Relation}
\label{sec:dynamics}
In this section we examine a dynamical argument, originally proposed by \citet{Wang2009}, for the presence of dual AGN in our sample.  Under the simplest picture of Keplerian orbital motion, a binary system of masses should show an inverse correlation between the ratio of their velocities, $V_{1}/V_{2}$, and the ratio of their masses, $M_{1}/M_{2}$, i.e. $V_{1}/V_{2}=M_{2}/M_{1}$.  As shown in \citet{Wang2009} for their sample of double-peaked \oiii~AGN, the ratio of the line offsets is equivalent to the velocity ratio, and the mass ratio can be approximated by the luminosity ratio multiplied by a factor representing the ratio of the accretion rates, $\epsilon_{1,2}$.  This yields the relation $L_{b}/L_{r}=\epsilon_{b,r}\Delta V_{r}/\Delta V_{b}$ between the blue and red components.  For our sample, we have measured the requisite observational properties to perform this same analysis, but with the emission lines \nev~and \neiii.

Figure \ref{fig:flux_shift} shows $\Delta V_{b}/\Delta V_{r}$ plotted against $L_{b}/L_{r}$ for both \nev~ and \neiii.  The plots include both our sample and the low-redshift comparison sample.  Also over-plotted are the same confirmed/strong dual AGN candidates and extended/unresolved NLR AGN shown in Figure \ref{fig:strat} and discussed in Section \ref{variation}.  From a Spearman rank correlation test, it is clear that there is a strong negative correlation between $\Delta V_{b}/\Delta V_{r}$ and $L_{b}/L_{r}$, which is in the same sense as expected based on the theoretical relation of the binary, Keplerian orbit.  Specifically, a strong correlation is seen individually in our sample, the comparison sample, and in the combined sample.  For both \nev~ and \neiii, the comparison sample shows slightly stronger correlations than for our sample.  

As argued in \citet{Wang2009}, in a major merger one might expect that the SMBHs will be in similar environments and therefore, on average, should have similar Eddington ratios ($\epsilon_{b,r}=1$).  Therefore, we over-plotted the theoretical relation for $\epsilon_{b,r}=1$, which has the following form in base-10 logarithm space: $\log_{10}[L_{b}/L_{r}]=\log_{10}[\epsilon_{b,r}]-\log_{10}[V_{b}/V_{r}]$.  For both \nev~ and \neiii~ the colored/starred sources are generally consistent with the theoretical and best-fit lines, and they span a large enough dynamic range to display the same general trend as our sample.    The best-fit linear relations (weighted by the X and Y errors) are $\log_{10}[L_{b}/L_{r}]=(0.16\pm0.02)-(0.75\pm0.09)\times \log_{10}[V_{b}/V_{r}]$ (\nev) and $\log_{10}[L_{b}/L_{r}]=(0.05\pm0.02)-(0.51\pm0.04)\times \log_{10}[V_{b}/V_{r}]$ (\neiii).  These relations are similar to the theoretical relation for $\epsilon_{b,r}=1$, though they are technically not consistent when accounting for the $1\sigma$ errors, with both fits yielding shallower slopes.  The shallower slopes appear to be caused by some of the sources at the low $L_{b}/L_{r}$ end which extend out to larger $V_{b}/V_{r}$ ratios than the rest of the sample.  In this portion of the plot, our sources are systematically shifted above the theoretical relation.  Finally, we have over-plotted the theoretical relation expected for a biconical outflow, $L_{b}/L_{r}=(\Delta V_{b}/\Delta V_{r})^{3}$ (developed in \citealt{Wang2009}).  This relation is in the nearly the opposite sense as the for the binary relation and is far from agreeing with the best-fit relations.  In Section \ref{sec:test_dyn} we will discuss how this correlation and the offset from the theoretical relation may be consistent with the presence of both dual AGN and outflows in our sample.

While the version of this analysis presented in \citet{Wang2009} utilized host galaxy redshifts (obtained through fitting template galaxy spectra), we can not obtain that information for our sample since the quasar continuum outshines the galaxy starlight.  Instead, the velocity offsets used in our analysis here are relative to the SDSS redshifts ($z_{SDSS}$) discussed in Section \ref{redshifts}.  However, we have attempted to investigate the dependence of our results on the choice of redshift.  Recalling our discussion in Section \ref{redshifts}, we also have redshifts based upon \mgii~ which (in the single SMBH scenario) should trace the central, active SMBH redshift.  Therefore, we examined $V_{b}/V_{r}$ versus $L_{b}/L_{r}$ for the subset of our sample with robust (percent error $<0.1\%$) $z_{MgII}$ values.  We find that this does not introduce a significant change in the best-fit coefficients for \nev~ or \neiii.

\section{Radio Loudness}
\label{radio}
Since $91\%$ (119/131) of our sources are in the FIRST footprint ($5\sigma$ flux limit of $\sim$750 mJy), to determine their radio-loudness we have adopted the commonly used definition of radio to optical luminosity ratio $\mathcal{R}=L${\scriptsize5GHz}$/L${\scriptsize2500~\AA}, with $\mathcal{R}=10$ being the cutoff value for the radio-loud classification \citep{Ho:2002}.  From our parent sample, the fraction of radio-loud SDSS quasars in the FIRST footprint is 10\%, whereas that fraction is 23\% in our final double-peaked \nevs/\neiiis~ sample.  For comparison, \citet{Smith:2010} obtained 9\% (parent sample at $z<0.8$) and 27\% (double-peaked \oiii~ sample), respectively.  This preferential selection of radio sources over the parent quasar population in both studies suggests that the origin of the double-peaks might be related to the presence of radio jets in some sources, as in the case of SDSS J151709.20+335324.7 \citep{Rosario:2010}.

\section{Interpretation}
\label{interpretation}
In this section, we synthesize the results of the previous sections to highlight the most likely physical scenarios that produce the line splitting and line offsets in our sample of quasars with high-ionization, double-peaked narrow emission lines.  In particular, we examine the possibilities of AGN outflows and dual AGN.

\subsection{Examining the Outflow Hypothesis}
\label{outflows}
\begin{figure}[t!] $
\begin{array}{cc}
\hspace*{-0.in}\includegraphics[width=3.5in,height=2.625in]{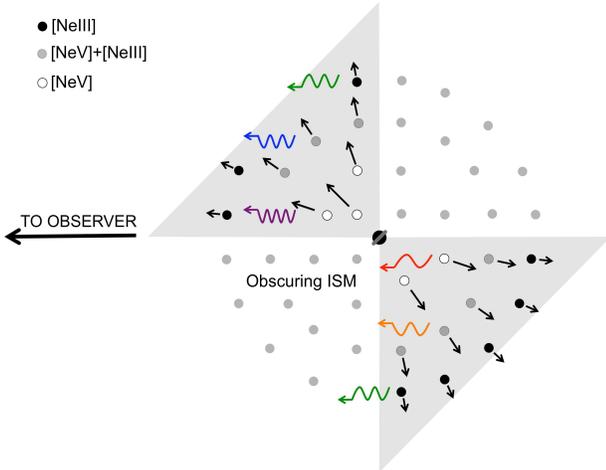}
\end{array} $
\caption{\footnotesize{Cross-section of the NLR showing our proposed geometry of the outflow/ionization stratification scenario and the relative origins of \nev~and \neiii.  The ISM around the central AGN is assumed to be spherically symmetric, the orientation of the outflow axis is $45^{\circ}$ from face-on, and we only show the NLR out to a radius which includes all of the \nev~and \neiii~emission.  The bi-cone of ionizing radiation (with an opening angle of $45^{\circ}$) emanating from the central SMBH is indicated by the gray shaded regions.  Various individual portions of the ISM and NLR are indicated by solid circles.  Portions where \nev~is produced but not \neiii~are shown as solid white circles, portions where both \nev~and \neiii~ are produced are indicated by solid gray circles, and portions where \neiii~is produced but not \nev~are indicated by solid black circles.  Portions of the ISM not in the ionization cone are indicated by light gray circles.  The black arrows indicate the velocities relative to the central SMBH, and the colored waves/arrows represent Doppler shifting of the light due to the velocity component along the line of sight to the observer.}}
\label{fig:cartoon}
\end{figure}

The correlation between line-splitting and line width evident in the blue and red systems of both \nev~ and \neiii~(Figure \ref{relv_width}) indicates that the mechanisms producing both the line-splittings and the line broadening are related, an observation which is consistent with the two emission line components originating near the same AGN.  In general, emission line velocity-offsets from the host galaxy or quasar redshift are often interpreted as evidence for outflowing photoionized gas; powerful AGN, including quasars, are known to be capable of driving high-velocity and/or large-scale outflows \citep{Fischer:2011}.  We will discuss how properties of our sample are consistent with some of the mechanisms known to produce outflows, and how they result in stratified NLRs.  Furthermore, we will discuss our interpretation within the context of a proposed outflow and stratification geometry shown in Figure \ref{fig:cartoon}.

\subsubsection{Mechanisms Producing AGN Outflows}
Two of the commonly proposed mechanisms for driving outflows in powerful AGN are radiation pressure from the accretion disk and radio jets.  Figure \ref{fig:ledd} shows that the velocity-splittings are correlated with the quasar Eddington ratio, $f_{\rm{Edd}}$, for \nev~ and for \neiii.  This result is consistent with the notion that outflows can be driven by radiation pressure from an accretion disk, and that more actively accreting SMBHs will drive stronger outflows.  This scenario is shown in Figure \ref{fig:cartoon} with the radiation emanating from the accretion disk in the commonly assumed bi-conical shape \citep{Antonucci:1993}.  Our result is also consistent with the results of \citet{Komossa:2008b} who find that their sample of NLS1s with offset \oiii~lines have relatively large Eddington ratios which might be reflective of the radiation driving the outflow.  Finally, that the correlation with $f_{\rm{Edd}}$ is stronger for \nev~ than for \neiii~ fits with a picture in which the NLR gas closest to the central engine is more strongly accelerated by the radiation pressure, as indicated in Figure \ref{fig:cartoon} by the larger velocity vectors on material closest to the SMBH.

However, it is possible that not all of the outflows are driven by radiation pressure.  In particular, radio jets are known to drive outflows in AGN by entrainment of NLR material, and our sample (and the \oiiis-selected sample) have radio loud fractions of $\sim$25\% (Section \ref{radio}).  Therefore, the high fraction of radio loud quasars in our sample relative to the parent quasar sample is a strong indication that the double-peaked sample we have compiled includes quasars with outflowing components.

\subsubsection{Evidence for Stratified NLRs}
If outflows are a common mechanism for driving line-splitting in NLRs, then this should naturally result in a stratified NLR since lines of greater I.P. and $n_{\rm{crit}}$ will be preferentially produced nearer to the AGN where they are accelerated to higher velocities relative to lines of lower I.P and $n_{\rm{crit}}$.  The relative origins and velocities of emission lines in the stratification scenario are illustrated in Figure \ref{fig:cartoon}, where \nev~originates closer to the SMBH than \neiii.  We examined this scenario by looking at the relationship between the velocity-offsets of \nev~ and \neiii~ in Section \ref{variation}.  In that analysis, Figure \ref{fig:strat}A shows that 60\% of the sources have a significantly larger blue velocity offset in \nev~compared to \neiii, suggestive of a stratified NLR.   We are only seeing the projected velocity-offsets, so this percentage might represent a lower limit on the number of sources with stratified NLRs when accounting for random orientations of the outflow axes.  We note that in Figure \ref{fig:strat} the largest stratifications are seen at the largest velocities.  This could also be a result of projection effects, since orientations which reduce the radial velocity components will also reduce the observed stratification.  However, this trend may also be due, in part, to the physical effect of stronger outflows (larger velocities) producing larger stratifications.

For most of the sources in our sample we can only place upper limits on the fluxes and velocity offsets of blue \oii~ components.  This is partly due to the blending of the $\lambda3726,3729$\AA~ doublet (Section \ref{sec:initial}).  However, if the double emission line components are driven by outflows and the NLR is stratified, one might expect that \oii, with a relatively small ionization potential of $I.P.=13.614$ eV and critical density of $n_{\rm{crit}}=3.4\times 10^{3}$ cm$^{-3}$, will be strongest in a portion of the NLR relatively further from the central AGN.  As a result, it will not be accelerated to high velocities, resulting in a small line-splitting.

With the outflow axis oriented at some angle intermediate to edge-on or face-on, the redshifted NLR emission will consequently be more attenuated than the blueshifted emission, as illustrated by the `obscuring ISM' labeled in Figure \ref{fig:cartoon}.  The portions of the NLR with the largest line of sight velocity components will be most obscured.  Conversely, the portions with the smallest line of sight velocity components will be the least obscured.  The result is that, in the presence of such attenuation, the most redshifted portion of the emission lines will obscured, moving the observed position of the red emission peak closer to the systemic, i.e. non-Doppler shifted, redshift.  In this case the red component resembles the `classical' NLR.  This is consistent with observations in which offset narrow emission lines in AGN are usually blueward of the systemic velocity, indicating that we are able to view the outflowing component moving toward the observer, while the component moving away from the observer is obscured by a larger column of dust.  For example, from Figure \ref{offsets} it appears that the mean magnitude of the red component velocity-offsets from the systemic redshift are generally smaller than those of the blue components, consistent with the notion that the red components are less dominated by outflows.

Note that in Figure \ref{fig:cartoon}, with sufficient attenuation even the NLR emission which is least Doppler-shifted will be obscured, resulting an apparent blueshifting of the red component, as is occasionally seen in some of our sources and in other studies \citep{Spoon:Holt:2009}.  For example, the blue \nev~component may be emitted from a portion of the NLR on the observer's side which is closest to the central source and moving at the greatest velocity (e.g. white dots in Figure \ref{fig:cartoon}), while the red \nev~component is from a portion which is further from the central source (e.g. grey dots in Figure \ref{fig:cartoon}).  We did find in Section \ref{ledd} evidence for a mild positive correlation between the red line offsets and the quasar Eddington ratio.  This suggests that, while the red system tends to represent the `classical' NLR, it is still effected by the radiation pressure since it must originate close enough to the central source where the ionizing flux is sufficient.  Additionally, Figures \ref{fig:strat}B and \ref{fig:strat}C show that there is some evidence for stratification of the red systems (though much less significant than for the blue systems).

\subsection{Implications for Dual AGN at High-Redshift}
It is possible that some of the sources in our sample may host two SMBHs following a galaxy merger.  In this case, the double-peaks may be from two distinct NLRs that each accompanies its own active SMBH, or perhaps two NLR peaks are influenced by the orbital motion of two SMBHs \citep{Blecha:2012}.  So far there are only a handful of known plausible merger remnants hosting two AGN at redshifts comparable to our sample: $z\sim0.709$ \citep{Gerke2007}, $z\sim0.78$ \citep{Comerford2009a}, and $z\sim1.175$ \citep{Barrows:2012}.  Since galaxy mergers were more frequent at higher redshifts, we would like to investigate the dual AGN scenario for the sources in our sample.

\subsubsection{Sources with No Apparent Ionization Stratification}
The two \oiiis-selected confirmed dual AGN for which we measured double peaks in \nev~and \neiii~are consistent with no apparent ionization stratification (Figure \ref{fig:strat}).  Additionally, there is only one candidate dual AGN identified through the double-peaked profile of \nev~and \neiii, and it is also consistent with no apparent ionization stratification \citep{Barrows:2012}.  In these sources, the evidence for outflowing NLR material is less compelling, and the line-splitting may instead be produced by orbital motion of two AGN about each other.  Likewise, the $40\%$ percent of our sources plotted in Figure \ref{fig:strat} with no apparent ionization stratification may include cases where the double-peaks are the result of two SMBHs following a galaxy merger.  Additionally, we see explicit double \oii~peaks for a subset (11) of our sources which perhaps suggests that the outflow scenario is less likely in these sources since \oii~should originate at a greater distance from the central AGN, as mentioned in Section \ref{outflows}.

We note, however, that the lack of an apparent ionization stratification does not preclude the possibility of an outflow, or general gas kinematic origin of the double-peaked emission lines.  For example, the two extended NLR AGN from \citet{Fu:2012} have no measurable ionization stratification but the double emission components are known to be produced by the NLR around a single AGN based on integral-field spectroscopy and high-resolution imaging.  Conversely, a stratified NLR (or two stratified NLRs) does not preclude the presence of two AGN.  For example, as discussed in Section \ref{variation}, our measurement of \nevs/\neiii~ in the confirmed dual AGN SDSS J150243.1+111557 shows some evidence for a stratification.
~\\
\subsubsection{Dual AGN with Large Velocities}
As is evident from Figure \ref{offsets} and discussed in Section \ref{sec:offsets}, the velocity-splittings in our sample are generally larger than those from \oiiis-selected samples, which might tend to select against likely dual AGN candidates at kpc-scale separations since they would not be bound to the merging galaxy system with such large velocities.  Most strong dual AGN candidates have \dv s less than $500$ km s$^{-1}$: $\Delta V=150$ km s$^{-1}$ \citep{Comerford2009b,Civano2010}, $\Delta V=500$ km s$^{-1}$ \citep{XK09}, $\Delta V=350$ km s$^{-1}$ \citep{Comerford:2011}, and $\Delta V=420$ km s$^{-1}$ \citep{McGurk:2011,Fu:2012}.  

However, there are several candidates with velocities $>500$ km s$^{-1}$: $\Delta V=630$ km s$^{-1}$ \citep{Gerke2007} and $\Delta V=700$ km s$^{-1}$ \citep{Barrows:2012}. 
Additionally, the dual AGN hypothesis could be allowed for larger \dv s if the AGN pairs are at small separations.  For example, \citet{Blecha:2012} find in their simulations that large \dv s ($>500$ km s$^{-1}$) are often associated with dual AGN at sub-kpc separations during pericentric passages.  For comparison, $23\%$ of our sample have $\Delta V<500$ km s$^{-1}$, $45\%$ have $500<\Delta V<800$ km s$^{-1}$, and $32\%$ have $\Delta V>800$ km s$^{-1}$, with a maximum of $\Delta V=1665$ km s$^{-1}$.  Therefore, though a fraction of our sample exhibit \dv s higher than expected for dual AGN, $2/3$ fall in the range expected for either kpc or sub-kpc separation AGN pairs.  We note that recent numerical simulations suggest dual activation of the SMBHs following a galaxy merger is most likely to occur at separations smaller than 1-10 kpc \citep{Van_Wassenhove:2012}.  Therefore, under this picture of dual activation, the correlation between Eddington ratio and line-splitting seen in Figure \ref{fig:ledd} would naturally emerge for a sample of dual AGN.

As discussed in Section \ref{redshifts}, for dual AGN with sufficiently large orbital velocities the broad emission line profiles may be significantly broader than expected if both components are Type 1 AGN.  To test for such additional broadening, we have compared the \mgii~FWHMs with those of \feii, but find no evidence for systematically broadened \mgii~compared to \feii.  Furthermore, a K-S test does not indicate a significant difference between the \mgii~FWHM distribution of our sample and that of the parent sample.  However, we note that the mean $FWHM_{\rm{MgII}}$ value for our sample (4740 km s$^{-1}$) is slightly larger than that of the parent sample (4580 km s$^{-1}$) which is perhaps suggestive of additional \mgii~broadening.

\subsubsection{Testing the Dynamical Argument}
\label{sec:test_dyn}
In Section \ref{sec:dynamics} we tested a dynamical argument for the presence of dual AGN in our sample.  The results are generally consistent with the theoretical expectation for a binary, Keplerian orbit (Figure \ref{fig:flux_shift}).  However, to understand the extent to which we can interpret this result, we must also strongly consider the role of alternative physical scenarios in producing such a correlation.

First, we note that for both \nev~and \neiii, based on the Spearman rank test, the comparison sample shows slightly stronger correlations than for our sample.  If this is an indication that the comparison sample shows stronger evidence for dual AGN, it would be consistent with the notion that larger velocity-splittings are less likely to be associated with dual AGN since our sample has larger velocity-splittings than the comparison sample.

Second, it is worth noting that the coefficients for the best-fit linear relations of \nev~and \neiii~are in disagreement.  This is consistent with the correlation being produced by outflows (or at least some of the sources experiencing outflows) since the red component of \neiii~would originate at a greater distance from the observer (compared to the red \nev~component) and therefore be even more obscured relative to the blue component (Figure \ref{fig:cartoon}).  In this case, the $L_{b}/L_{r}$ ratio should be even larger for \neiii, making the slope shallower as observed. 

However, it is possible that if our sample contains some combination of outflows and dual AGN then the outflows are responsible for deviations from the theoretical binary relation.  For example, as discussed in Section \ref{sec:dynamics} many of the sources at the low $L_{b}/L_{r}$, high $V_{b}/V_{r}$ portion of Figure \ref{fig:flux_shift} are offset above the theoretical relation.  These sources may be more likely to represent outflows since they are trending in the same direction as the outflow relation.  Additionally, they have large $V_{b}/V_{r}$ ratios because the red component is near the systemic redshift, consistent with attenuation of the redshifted outflow component.  Lastly, the $L_{b}/L_{r}$ ratios are smaller, indicating that the red component is stronger, and the blue component is a lower luminosity, extended wing as is often seen in outflows and is seen a few of our sources.  If these deviant sources are most likely to be outflows, then the remainder would be more consistent with the theoretical relation.  Additionally, the remainder would have a $L_{b}/L_{r}$ distribution consistent with $\epsilon_{b,r}=1$, similar to the result of \citet{Wang2009}.

We note that an additional source of scatter in the correlation could be due to stochastic accretion, such that the luminosity ratio does not accurately reflect the true mass ratio.  This effect could be particularly significant when the SMBHs are at larger separations when gas is less efficiently funneled to the nuclear regions.  Interestingly, at $z=0.8-1.6$, the 3'' fiber diameter of the SDSS spectrograph corresponds to $\sim$22-25 kpc, so that our sample may contain such early-stage mergers.

\subsubsection{Estimating the Fraction of Dual AGN in Our Sample}
While the completeness of our selection process as discussed in Section \ref{sec:offsets} suggests that there is a significant number of double-peaked emitters that we have missed, especially at small \dv s, we can not correct for the true number since we do not know the shape of the underlying distribution.  However, at the least we can use our \oiiis-selected comparison sample to estimate what fraction of \oiii~double-peaked emitters that we missed based on selection through \nev~or \neiii.  Of the 57 Type 1 AGN from \citep{Smith:2010}, we could reliably measure double \neiii~peaks for $67\%$ (we note that this fraction is consistent with the comparison between our completeness estimates and those of \citet{Liu2010a} in Section \ref{completeness}).  Based on this fraction, we arrive at a corrected number of 195 double-peaked \oiii~sources and a double-peaked \oiii~AGN fraction of $\sim0.5\%$ at $0.8<z<1.6$.  This fraction is likely to be a lower estimate because the double-peaked \nev~and \neiii~lines of the \oiiis-selected sample were not selected in exactly the same way (i.e. we had a prior knowledge of the double-peaked separation).  Therefore, the double-peaked \oiii~AGN fraction at $0.8<z<1.6$ is likely to be larger, potentially making it consistent with the fraction of $\sim1\%$ found by \citet{Liu2010a} at $z<0.8$.

To actually estimate the expected fraction of dual AGN in our sample we need the true fractions of double-peaked AGN and of dual AGN out of all AGN at $0.8<z<1.6$, neither of which are known.  However, we may make several reasonable assumptions that provide a rough estimate.  First, in order to determine the true fraction of double-peaked AGN out of all AGN at $0.8<z<1.6$, we need to correct for both our selection incompleteness and random projection effects.  This fraction was estimated by \citet{Shen:2011b} in which they determined that the fraction of detectable double-peaked \oiii~AGN is only $20\%$ of the actual number of AGN with double \oiii~components.  Correcting our estimated double-peaked \oiii~fraction of $0.5\%$ yields a `true' double \oiii~fraction of $\sim$2.5\% at $0.8<z<1.6$.  The influence of inclination and phase angle for selection of dual AGN through double-peaked emission line profiles was also investigated by \citet{Wang:2012} in which they found that, at a phase angle of $\phi\approx 50^{\circ}$, we miss at least 50\% of all AGN with double emission components.  We note that this 50\% correction is a lower limit since it does not account for instrumental resolution, and that applying additional corrections based on our completeness estimates would likely make the correction estimated in this manner similar to that of \citet{Shen:2011b}.  Finally, if we take the dual AGN fraction at $z=1.2$ to be $\sim0.05\%$ \citep{Yu:2011}, then we estimate the fraction of dual AGN out of double-peaked AGN at $0.8<z<1.6$ to be $2\%$.  This fraction is several times smaller than the results of \citet{Fu:2012} (4.5-12\%) and \citet{Shen:2011b} ($\sim10\%$) which were obtained from follow-up observations of double-peaked \oiii~AGN.  However, the difference can be attributed, at least in part, to the small expected number of dual AGN at $z=1.2$ estimated by \citet{Yu:2011} which is due to the redshift evolution of galaxy morphology in their analysis which yields more late-type galaxies with smaller initial SMBH masses at higher redshift.  

A direct test of this through follow-up observations is therefore crucial in understanding the frequency of galaxy mergers at redshifts $z>0.8$ and their role in AGN triggering.  For example, NIR spectroscopy will be capable of accessing the redshifted \oiii~emission line for sources in our sample, allowing for a direct comparison with the $z<0.8$ samples of double-peaked AGN.  This was done with the $z=1.175$ dual AGN candidate CXOXBJ142607.6+353351 in \citet{Barrows:2012} which was initially selected through double-peaked \nevs/\neiiis~but for which follow-up NIR spectroscopy provided access to \oiii.  The additional spatial information of \oiii~provided by 2D longslit spectroscopy would enable one to determine if any of these sources are strong dual AGN candidates.  Follow-up high-resolution imaging, such as radio observations, would be capable of resolving the two AGN cores, if present.

\section{Conclusions}
\label{conclusions}
We have compiled a sample of 131 quasars at $z=0.8-1.6$ which show double emission line components in either of the high-ionization narrow lines \nev~and \neiii.  The purpose of this search was to identify high-redshift analogs of the double-peaked \oiii~sources found in several previous studies.  Those double-peaked \oiii~sources are believed to represent complex gas kinematics, large-scale outflows, or in a few cases dual AGN.  Given the increased frequency of galaxy mergers at higher redshifts and their importance in models of galaxy evolution, we have investigated these phenomena at higher redshifts using our sample, with the following conclusions:

\begin{itemize}

\item[$\circ$] \hspace*{-0.05in} There is a clear bias towards selecting double-peaks with large velocity-splittings. This bias was made apparent by our comparison of the velocity-offsets of the blue and red components in our sample to those of \oiiis-selected samples, and it is corroborated by the results of our completeness simulations.  This selection bias is not surprising, and it is imposed by  the relatively weaker intensities of \nev~and \neiii~compared to \oiii.

\item[$\circ$] \hspace*{-0.05in} We have found two results suggesting that both the blue and red systems are influenced by kinematics in the NLR. First, the line-widths of both the blue and red components are strongly correlated with the line-splittings, suggesting a common origin.  Second, we find that the individual offsets for both the `blue' and `red' systems are positively correlated with the quasar Eddington ratio, suggesting that the SMBH accretion rate and therefore the radiation pressure is responsible for driving the line-offsets in the blueward direction for both line components.  

\item[$\circ$] \hspace*{-0.05in} We find evidence suggesting that the observed kinematics are strongest in the blue systems.  This is suggested because the blue systems' have larger velocity shifts from the quasar redshift, those velocity offsets show the strongest correlations with the Eddington ratio, and the blue systems show the highest degree of ionization stratification.  This further suggests that the red outflowing components are generally more obscured.

\item[$\circ$] \hspace*{-0.05in} We have found that a significant fraction ($\sim23\%$) are radio loud, compared to the $10\%$ radio loud fraction of the parent sample.

\end{itemize}

Taken together, the previous conclusions paint a picture in which the blue systems originate in a portion of the NLR much closer to the AGN where they are accelerated by the accretion disk radiation pressure or radio jets to high velocities.  This explains the large blueshifts, the stronger correlation with Eddington ratio and the pronounced ionization stratification.  The red system originates further from the AGN where it is not accelerated to the high velocities of the blue system but is nevertheless close enough so it's bulk velocity offset is also influenced by the AGN radiation pressure.  This explains the smaller velocity offsets, the much weaker correlation with Eddington ratio, and the relatively less pronounced ionization stratification compared to the blue systems.  A generalized schematic of this scenario in Figure \ref{fig:cartoon}.  The enhanced radio loud fraction relative to the parent sample also suggests that radio jets may be another mechanism which is capable of accelerating the NLR clouds to produce the line offsets.  This sample can be used to study outflows from luminous AGN at relatively high redshifts when AGN feedback may have been an important factor in the growth of massive galaxies. 

There are several interesting results which leave open the possibility for dual AGN in the sample.  In particular, several of our correlations can be thought of as consistent with the dual AGN scenario, and even suggest that the sample is likely to include dual AGN: \\

\begin{itemize}

\item[$\circ$] \hspace*{-0.05in} The correlation between velocity-splitting and Eddington ratio, while consistent with the picture of radiatively-driven outflows, could plausibly be consistent with orbiting SMBH pairs in which enhanced accretion is more likely to occur at smaller separations where the SMBH orbital velocities will be largest. 

\item[$\circ$] \hspace*{-0.05in} We have found that a subset of our sample ($40\%$) are consistent with no measurable ionization stratification between \nev~ and \neiii, similar to other dual AGN and strong candidate dual AGN.

\item[$\circ$] \hspace*{-0.05in} We have found that our sample shows a correlation between the velocity-offset ratio and the luminosity ratio of the blue and red components.  This correlation is broadly consistent with the theoretical expectation for a binary, Keplerian orbit, though our sample seems to be systematically offset above the relation.  We have shown how this deviation could be produced in a sample which includes a combination of AGN outflows and dual AGN. \\

\item[$\circ$] \hspace*{-0.05in} We have estimated the fraction of dual AGN out of double-peaked AGN that we expect at $0.8<z<1.6$, finding a fraction (2\%) which is smaller than that estimated at lower redshifts.  However, we caution that a significant - and perhaps the primary - reason for this lower fraction is the small estimated number of high redshift dual AGN that we adopt.  

\end{itemize}

Follow-up NIR observations to access the \oiii~ line in our sources would allow for a direct comparison with the \oiii~ velocity and spatial profiles of the $z<0.80$ samples, allowing for a more robust assessment of the origin of the double-peaks in the high-ionization narrow emission lines.  Therefore, this sample represents an initial step toward extending the study of double-peaked emission line AGN to higher redshifts.

We would like to thank an anonymous referee for very helpful comments that improved the quality of the paper.  We would also like to acknowledge Daniel Stern for assistance on the analysis of our data and Laura Blecha for highly useful suggestions regarding the interpretation of our results.  Finally, we acknowledge constructive discussion from members of the Arkansas Galaxy Evolution Survey (AGES) and the Arkansas Center for Space and Planetary Sciences, including Douglas Shields, Benjamin Davis, Adam Hughes and Jennifer Hanley.  This research has made use of NASA's Astrophysics Data System and the Sloan Digital Sky Survey.  Support for this work was provided in part by the Arkansas NASA EPSCoR program (grant number NNX08AW03A). 
~\\


\begin{thebibliography}{93}
\expandafter\ifx\csname natexlab\endcsname\relax\def\natexlab#1{#1}\fi

\bibitem[{{Abazajian} {et~al.}(2009){Abazajian}, {Adelman-McCarthy},
  {Ag{\"u}eros}, {Allam}, {Allende Prieto}, {An}, {Anderson}, {Anderson},
  {Annis}, {Bahcall}, {Bailer-Jones}, {Barentine}, {Bassett}, {Becker},
  {Beers}, {Bell}, {Belokurov}, {Berlind}, {Berman}, {Bernardi}, {Bickerton},
  {Bizyaev}, {Blakeslee}, {Blanton}, {Bochanski}, {Boroski}, {Brewington},
  {Brinchmann}, {Brinkmann}, {Brunner}, {Budav{\'a}ri}, {Carey}, {Carliles},
  {Carr}, {Castander}, {Cinabro}, {Connolly}, {Csabai}, {Cunha}, {Czarapata},
  {Davenport}, {de Haas}, {Dilday}, {Doi}, {Eisenstein}, {Evans}, {Evans},
  {Fan}, {Friedman}, {Frieman}, {Fukugita}, {G{\"a}nsicke}, {Gates},
  {Gillespie}, {Gilmore}, {Gonzalez}, {Gonzalez}, {Grebel}, {Gunn},
  {Gy{\"o}ry}, {Hall}, {Harding}, {Harris}, {Harvanek}, {Hawley}, {Hayes},
  {Heckman}, {Hendry}, {Hennessy}, {Hindsley}, {Hoblitt}, {Hogan}, {Hogg},
  {Holtzman}, {Hyde}, {Ichikawa}, {Ichikawa}, {Im}, {Ivezi{\'c}}, {Jester},
  {Jiang}, {Johnson}, {Jorgensen}, {Juri{\'c}}, {Kent}, {Kessler}, {Kleinman},
  {Knapp}, {Konishi}, {Kron}, {Krzesinski}, {Kuropatkin}, {Lampeitl},
  {Lebedeva}, {Lee}, {Lee}, {Leger}, {L{\'e}pine}, {Li}, {Lima}, {Lin}, {Long},
  {Loomis}, {Loveday}, {Lupton}, {Magnier}, {Malanushenko}, {Malanushenko},
  {Mandelbaum}, {Margon}, {Marriner}, {Mart{\'{\i}}nez-Delgado}, {Matsubara},
  {McGehee}, {McKay}, {Meiksin}, {Morrison}, {Mullally}, {Munn}, {Murphy},
  {Nash}, {Nebot}, {Neilsen}, {Newberg}, {Newman}, {Nichol}, {Nicinski},
  {Nieto-Santisteban}, {Nitta}, {Okamura}, {Oravetz}, {Ostriker}, {Owen},
  {Padmanabhan}, {Pan}, {Park}, {Pauls}, {Peoples}, {Percival}, {Pier}, {Pope},
  {Pourbaix}, {Price}, {Purger}, {Quinn}, {Raddick}, {Fiorentin}, {Richards},
  {Richmond}, {Riess}, {Rix}, {Rockosi}, {Sako}, {Schlegel}, {Schneider},
  {Scholz}, {Schreiber}, {Schwope}, {Seljak}, {Sesar}, {Sheldon}, {Shimasaku},
  {Sibley}, {Simmons}, {Sivarani}, {Smith}, {Smith}, {Smol{\v c}i{\'c}},
  {Snedden}, {Stebbins}, {Steinmetz}, {Stoughton}, {Strauss}, {Subba Rao},
  {Suto}, {Szalay}, {Szapudi}, {Szkody}, {Tanaka}, {Tegmark}, {Teodoro},
  {Thakar}, {Tremonti}, {Tucker}, {Uomoto}, {Vanden Berk}, {Vandenberg},
  {Vidrih}, {Vogeley}, {Voges}, {Vogt}, {Wadadekar}, {Watters}, {Weinberg},
  {West}, {White}, {Wilhite}, {Wonders}, {Yanny}, {Yocum}, {York}, {Zehavi},
  {Zibetti}, \& {Zucker}}]{Abazajian09}
{Abazajian}, K.~N., {et~al.} 2009, ApJ, 182, 543

\bibitem[{{Antonucci}(1993)}]{Antonucci:1993}
{Antonucci}, R. 1993, \araa, 31, 473

\bibitem[{{Arribas} {et~al.}(1996){Arribas}, {Mediavilla}, \&
  {Garcia-Lorenzo}}]{Arribas:1996}
{Arribas}, S., {Mediavilla}, E., \& {Garcia-Lorenzo}, B. 1996, \apj, 463, 509

\bibitem[{{Barrows} {et~al.}(2012){Barrows}, {Stern}, {Madsen}, {Harrison},
  {Assef}, {Comerford}, {Cushing}, {Fassnacht}, {Gonzalez}, {Griffith},
  {Hickox}, {Kirkpatrick}, \& {Lagattuta}}]{Barrows:2012}
{Barrows}, R.~S., {et~al.} 2012, \apj, 744, 7

\bibitem[{{Bennert} {et~al.}(2002){Bennert}, {Falcke}, {Schulz}, {Wilson}, \&
  {Wills}}]{Bennert:2002}
{Bennert}, N., {Falcke}, H., {Schulz}, H., {Wilson}, A.~S., \& {Wills}, B.~J.
  2002, \apjl, 574, L105

\bibitem[{{Berrier} {et~al.}(2006){Berrier}, {Bullock}, {Barton}, {Guenther},
  {Zentner}, \& {Wechsler}}]{Berrier:2006}
{Berrier}, J.~C., {Bullock}, J.~S., {Barton}, E.~J., {Guenther}, H.~D.,
  {Zentner}, A.~R., \& {Wechsler}, R.~H. 2006, \apj, 652, 56

\bibitem[{{Berrier} \& {Cooke}(2012)}]{Berrier:2012}
{Berrier}, J.~C., \& {Cooke}, J. 2012, ArXiv e-prints

\bibitem[{{Bianchi} {et~al.}(2008){Bianchi}, {Chiaberge}, {Piconcelli},
  {Guainazzi}, \& {Matt}}]{Bianchi2008}
{Bianchi}, S., {Chiaberge}, M., {Piconcelli}, E., {Guainazzi}, M., \& {Matt},
  G. 2008, \mnras, 386, 105

\bibitem[{{Blecha} {et~al.}(2012){Blecha}, {Loeb}, \& {Narayan}}]{Blecha:2012}
{Blecha}, L., {Loeb}, A., \& {Narayan}, R. 2012, ArXiv e-prints

\bibitem[{{Cisternas} {et~al.}(2011){Cisternas}, {Jahnke}, {Inskip},
  {Kartaltepe}, {Koekemoer}, {Lisker}, {Robaina}, {Scodeggio}, {Sheth},
  {Trump}, {Andrae}, {Miyaji}, {Lusso}, {Brusa}, {Capak}, {Cappelluti},
  {Civano}, {Ilbert}, {Impey}, {Leauthaud}, {Lilly}, {Salvato}, {Scoville}, \&
  {Taniguchi}}]{Cisternas:2011}
{Cisternas}, M., {et~al.} 2011, \apj, 726, 57

\bibitem[{{Civano} {et~al.}(2010){Civano}, {Elvis}, {Lanzuisi}, {Jahnke},
  {Zamorani}, {Blecha}, {Bongiorno}, {Brusa}, {Comastri}, {Hao}, {Leauthaud},
  {Loeb}, {Mainieri}, {Piconcelli}, {Salvato}, {Scoville}, {Trump}, {Vignali},
  {Aldcroft}, {Bolzonella}, {Bressert}, {Finoguenov}, {Fruscione}, {Koekemoer},
  {Cappelluti}, {Fiore}, {Giodini}, {Gilli}, {Impey}, {Lilly}, {Lusso},
  {Puccetti}, {Silverman}, {Aussel}, {Capak}, {Frayer}, {Le Floch},
  {McCracken}, {Sanders}, {Schiminovich}, \& {Taniguchi}}]{Civano2010}
{Civano}, F., {et~al.} 2010, \apj, 717, 209

\bibitem[{{Civano} {et~al.}(2012){Civano}, {Elvis}, {Lanzuisi}, {Aldcroft},
  {Trichas}, {Bongiorno}, {Brusa}, {Blecha}, {Comastri}, {Loeb}, {Salvato},
  {Fruscione}, {Koekemoer}, {Komossa}, {Gilli}, {Mainieri}, {Piconcelli}, \&
  {Vignali}}]{Civano:2012}
---. 2012, \apj, 752, 49

\bibitem[{{Comerford} {et~al.}(2012){Comerford}, {Gerke}, {Stern}, {Cooper},
  {Weiner}, {Newman}, {Madsen}, \& {Barrows}}]{Comerford:2012}
{Comerford}, J.~M., {Gerke}, B.~F., {Stern}, D., {Cooper}, M.~C., {Weiner},
  B.~J., {Newman}, J.~A., {Madsen}, K., \& {Barrows}, R.~S. 2012, \apj, 753, 42

\bibitem[{{Comerford} {et~al.}(2009{\natexlab{a}}){Comerford}, {Griffith},
  {Gerke}, {Cooper}, {Newman}, {Davis}, \& {Stern}}]{Comerford2009b}
{Comerford}, J.~M., {Griffith}, R.~L., {Gerke}, B.~F., {Cooper}, M.~C.,
  {Newman}, J.~A., {Davis}, M., \& {Stern}, D. 2009{\natexlab{a}}, \apjl, 702,
  L82

\bibitem[{{Comerford} {et~al.}(2011){Comerford}, {Pooley}, {Gerke}, \&
  {Madejski}}]{Comerford:2011}
{Comerford}, J.~M., {Pooley}, D., {Gerke}, B.~F., \& {Madejski}, G.~M. 2011,
  \apjl, 737, L19+

\bibitem[{{Comerford} {et~al.}(2009{\natexlab{b}}){Comerford}, {Gerke},
  {Newman}, {Davis}, {Yan}, {Cooper}, {Faber}, {Koo}, {Coil}, {Rosario}, \&
  {Dutton}}]{Comerford2009a}
{Comerford}, J.~M., {et~al.} 2009{\natexlab{b}}, \apj, 698, 956

\bibitem[{{Conselice} {et~al.}(2003){Conselice}, {Bershady}, {Dickinson}, \&
  {Papovich}}]{Conselice:2003}
{Conselice}, C.~J., {Bershady}, M.~A., {Dickinson}, M., \& {Papovich}, C. 2003,
  \aj, 126, 1183

\bibitem[{{Crenshaw} {et~al.}(2010{\natexlab{a}}){Crenshaw}, {Kraemer},
  {Schmitt}, {Jaff{\'e}}, {Deo}, {Collins}, \& {Fischer}}]{Crenshaw:2010b}
{Crenshaw}, D.~M., {Kraemer}, S.~B., {Schmitt}, H.~R., {Jaff{\'e}}, Y.~L.,
  {Deo}, R.~P., {Collins}, N.~R., \& {Fischer}, T.~C. 2010{\natexlab{a}}, \aj,
  139, 871

\bibitem[{{Crenshaw} {et~al.}(2010{\natexlab{b}}){Crenshaw}, {Schmitt},
  {Kraemer}, {Mushotzky}, \& {Dunn}}]{Crenshaw:2010a}
{Crenshaw}, D.~M., {Schmitt}, H.~R., {Kraemer}, S.~B., {Mushotzky}, R.~F., \&
  {Dunn}, J.~P. 2010{\natexlab{b}}, \apj, 708, 419

\bibitem[{{De Robertis} \& {Osterbrock}(1984)}]{De_Robertis:1984}
{De Robertis}, M.~M., \& {Osterbrock}, D.~E. 1984, \apj, 286, 171

\bibitem[{{Di Matteo} {et~al.}(2005){Di Matteo}, {Springel}, \&
  {Hernquist}}]{Di_Matteo:2005}
{Di Matteo}, T., {Springel}, V., \& {Hernquist}, L. 2005, \nat, 433, 604

\bibitem[{{Eracleous} \& {Halpern}(2003)}]{EH03}
{Eracleous}, M., \& {Halpern}, J.~P. 2003, ApJ, 599, 886

\bibitem[{{Everett}(2007)}]{Everett2007b}
{Everett}, J.~E. 2007, \apss, 311, 269

\bibitem[{{Everett} \& {Murray}(2007)}]{Everett2007a}
{Everett}, J.~E., \& {Murray}, N. 2007, \apj, 656, 93

\bibitem[{{Ferland} \& {Osterbrock}(1986)}]{Ferland:Osterbrock:1986}
{Ferland}, G.~J., \& {Osterbrock}, D.~E. 1986, \apj, 300, 658

\bibitem[{{Ferrarese} \& {Merritt}(2000)}]{Ferrarese:Merritt:2000}
{Ferrarese}, L., \& {Merritt}, D. 2000, \apjl, 539, L9

\bibitem[{{Fischer} {et~al.}(2011){Fischer}, {Crenshaw}, {Kraemer}, {Schmitt},
  {Mushotsky}, \& {Dunn}}]{Fischer:2011}
{Fischer}, T.~C., {Crenshaw}, D.~M., {Kraemer}, S.~B., {Schmitt}, H.~R.,
  {Mushotsky}, R.~F., \& {Dunn}, J.~P. 2011, \apj, 727, 71

\bibitem[{{Fu} {et~al.}(2011{\natexlab{a}}){Fu}, {Myers}, {Djorgovski}, \&
  {Yan}}]{Fu:2011a}
{Fu}, H., {Myers}, A.~D., {Djorgovski}, S.~G., \& {Yan}, L. 2011{\natexlab{a}},
  \apj, 733, 103

\bibitem[{{Fu} {et~al.}(2012){Fu}, {Yan}, {Myers}, {Stockton}, {Djorgovski},
  {Aldering}, \& {Rich}}]{Fu:2012}
{Fu}, H., {Yan}, L., {Myers}, A.~D., {Stockton}, A., {Djorgovski}, S.~G.,
  {Aldering}, G., \& {Rich}, J.~A. 2012, \apj, 745, 67

\bibitem[{{Fu} {et~al.}(2011{\natexlab{b}}){Fu}, {Zhang}, {Assef}, {Stockton},
  {Myers}, {Yan}, {Djorgovski}, {Wrobel}, \& {Riechers}}]{Fu:2011}
{Fu}, H., {et~al.} 2011{\natexlab{b}}, \apjl, 740, L44

\bibitem[{{Ge} {et~al.}(2012){Ge}, {Hu}, {Wang}, {Bai}, \& {Zhang}}]{Ge:2012}
{Ge}, J.-Q., {Hu}, C., {Wang}, J.-M., {Bai}, J.-M., \& {Zhang}, S. 2012, \apjs,
  201, 31

\bibitem[{{Gerke} {et~al.}(2007){Gerke}, {Newman}, {Lotz}, {Yan}, {Barmby},
  {Coil}, {Conselice}, {Ivison}, {Lin}, {Koo}, {Nandra}, {Salim}, {Small},
  {Weiner}, {Cooper}, {Davis}, {Faber}, \& {Guhathakurta}}]{Gerke2007}
{Gerke}, B.~F., {et~al.} 2007, \apjl, 660, L23

\bibitem[{{Green} {et~al.}(2010){Green}, {Myers}, {Barkhouse}, {Mulchaey},
  {Bennert}, {Cox}, \& {Aldcroft}}]{Green2010}
{Green}, P.~J., {Myers}, A.~D., {Barkhouse}, W.~A., {Mulchaey}, J.~S.,
  {Bennert}, V.~N., {Cox}, T.~J., \& {Aldcroft}, T.~L. 2010, \apj, 710, 1578

\bibitem[{{Guainazzi} {et~al.}(2005){Guainazzi}, {Piconcelli},
  {Jim{\'e}nez-Bail{\'o}n}, \& {Matt}}]{Guainazzi2005}
{Guainazzi}, M., {Piconcelli}, E., {Jim{\'e}nez-Bail{\'o}n}, E., \& {Matt}, G.
  2005, \aap, 429, L9

\bibitem[{{Heckman} {et~al.}(1981){Heckman}, {Miley}, {van Breugel}, \&
  {Butcher}}]{Heckman1981}
{Heckman}, T.~M., {Miley}, G.~K., {van Breugel}, W.~J.~M., \& {Butcher}, H.~R.
  1981, \apj, 247, 403

\bibitem[{{Hewett} \& {Wild}(2010)}]{Hewett:2010}
{Hewett}, P.~C., \& {Wild}, V. 2010, \mnras, 405, 2302

\bibitem[{{Ho}(2002)}]{Ho:2002}
{Ho}, L.~C. 2002, \apj, 564, 120

\bibitem[{{Holt} {et~al.}(2003){Holt}, {Tadhunter}, \& {Morganti}}]{Holt2003}
{Holt}, J., {Tadhunter}, C.~N., \& {Morganti}, R. 2003, \mnras, 342, 227

\bibitem[{{Holt} {et~al.}(2008){Holt}, {Tadhunter}, \& {Morganti}}]{Holt2008}
---. 2008, \mnras, 387, 639

\bibitem[{{Hopkins} {et~al.}(2005){Hopkins}, {Hernquist}, {Cox}, {Di Matteo},
  {Martini}, {Robertson}, \& {Springel}}]{Hopkins05}
{Hopkins}, P.~F., {Hernquist}, L., {Cox}, T.~J., {Di Matteo}, T., {Martini},
  P., {Robertson}, B., \& {Springel}, V. 2005, ApJ, 630, 705

\bibitem[{{Hudson} {et~al.}(2006){Hudson}, {Reiprich}, {Clarke}, \&
  {Sarazin}}]{Hudson2006}
{Hudson}, D.~S., {Reiprich}, T.~H., {Clarke}, T.~E., \& {Sarazin}, C.~L. 2006,
  \aap, 453, 433

\bibitem[{{Kauffmann} \& {Haehnelt}(2000)}]{Kauffman:2000}
{Kauffmann}, G., \& {Haehnelt}, M. 2000, \mnras, 311, 576

\bibitem[{{Kocevski} {et~al.}(2012){Kocevski}, {Faber}, {Mozena}, {Koekemoer},
  {Nandra}, {Rangel}, {Laird}, {Brusa}, {Wuyts}, {Trump}, {Koo}, {Somerville},
  {Bell}, {Lotz}, {Alexander}, {Bournaud}, {Conselice}, {Dahlen}, {Dekel},
  {Donley}, {Dunlop}, {Finoguenov}, {Georgakakis}, {Giavalisco}, {Guo},
  {Grogin}, {Hathi}, {Juneau}, {Kartaltepe}, {Lucas}, {McGrath}, {McIntosh},
  {Mobasher}, {Robaina}, {Rosario}, {Straughn}, {van der Wel}, \&
  {Villforth}}]{Kocevski:2012}
{Kocevski}, D.~D., {et~al.} 2012, \apj, 744, 148

\bibitem[{{Komatsu} {et~al.}(2011){Komatsu}, {Smith}, {Dunkley}, {Bennett},
  {Gold}, {Hinshaw}, {Jarosik}, {Larson}, {Nolta}, {Page}, {Spergel},
  {Halpern}, {Hill}, {Kogut}, {Limon}, {Meyer}, {Odegard}, {Tucker}, {Weiland},
  {Wollack}, \& {Wright}}]{Komatsu:2011}
{Komatsu}, E., {et~al.} 2011, \apjs, 192, 18

\bibitem[{{Komossa} {et~al.}(2003){Komossa}, {Burwitz}, {Hasinger}, {Predehl},
  {Kaastra}, \& {Ikebe}}]{Komossa2003}
{Komossa}, S., {Burwitz}, V., {Hasinger}, G., {Predehl}, P., {Kaastra}, J.~S.,
  \& {Ikebe}, Y. 2003, \apjl, 582, L15

\bibitem[{{Komossa} {et~al.}(2008){Komossa}, {Xu}, {Zhou}, {Storchi-Bergmann},
  \& {Binette}}]{Komossa:2008b}
{Komossa}, S., {Xu}, D., {Zhou}, H., {Storchi-Bergmann}, T., \& {Binette}, L.
  2008, \apj, 680, 926

\bibitem[{{Koss} {et~al.}(2011){Koss}, {Mushotzky}, {Treister}, {Veilleux},
  {Vasudevan}, {Miller}, {Sanders}, {Schawinski}, \& {Trippe}}]{Koss:2011}
{Koss}, M., {et~al.} 2011, \apjl, 735, L42+

\bibitem[{{Kriss}(1994)}]{Kriss94}
{Kriss}, G. 1994, Astronomical Data Analysis Software and Systems, 3, 437

\bibitem[{{Krolik}(1999)}]{Krolik:1999}
{Krolik}, J.~H. 1999, {Active galactic nuclei : from the central black hole to
  the galactic environment}, ed. {Krolik, J.~H.}

\bibitem[{{Liu} {et~al.}(2012{\natexlab{a}}){Liu}, {Civano}, {Shen}, {Green},
  {Greene}, \& {Strauss}}]{Liu:2012}
{Liu}, X., {Civano}, F., {Shen}, Y., {Green}, P.~J., {Greene}, J.~E., \&
  {Strauss}, M.~A. 2012{\natexlab{a}}, ArXiv e-prints

\bibitem[{{Liu} {et~al.}(2010{\natexlab{a}}){Liu}, {Greene}, {Shen}, \&
  {Strauss}}]{Liu2010b}
{Liu}, X., {Greene}, J.~E., {Shen}, Y., \& {Strauss}, M.~A. 2010{\natexlab{a}},
  \apjl, 715, L30

\bibitem[{{Liu} {et~al.}(2012{\natexlab{b}}){Liu}, {Shen}, \&
  {Strauss}}]{Liu:2012a}
{Liu}, X., {Shen}, Y., \& {Strauss}, M.~A. 2012{\natexlab{b}}, \apj, 745, 94

\bibitem[{{Liu} {et~al.}(2010{\natexlab{b}}){Liu}, {Shen}, {Strauss}, \&
  {Greene}}]{Liu2010a}
{Liu}, X., {Shen}, Y., {Strauss}, M.~A., \& {Greene}, J.~E. 2010{\natexlab{b}},
  \apj, 708, 427

\bibitem[{{Liu} {et~al.}(2011){Liu}, {Shen}, {Strauss}, \& {Hao}}]{Liu:2011}
{Liu}, X., {Shen}, Y., {Strauss}, M.~A., \& {Hao}, L. 2011, \apj, 737, 101

\bibitem[{{Lotz} {et~al.}(2011){Lotz}, {Jonsson}, {Cox}, {Croton}, {Primack},
  {Somerville}, \& {Stewart}}]{Lotz:2011}
{Lotz}, J.~M., {Jonsson}, P., {Cox}, T.~J., {Croton}, D., {Primack}, J.~R.,
  {Somerville}, R.~S., \& {Stewart}, K. 2011, \apj, 742, 103

\bibitem[{{Marconi} \& {Hunt}(2003)}]{Marconi:Hunt:2003}
{Marconi}, A., \& {Hunt}, L.~K. 2003, \apjl, 589, L21

\bibitem[{{Mazzarella} {et~al.}(2012){Mazzarella}, {Iwasawa}, {Vavilkin},
  {Armus}, {Kim}, {Bothun}, {Evans}, {Spoon}, {Haan}, {Howell}, {Lord},
  {Marshall}, {Ishida}, {Xu}, {Petric}, {Sanders}, {Surace}, {Appleton},
  {Chan}, {Frayer}, {Inami}, {Khachikian}, {Madore}, {Privon}, {Sturm}, {U}, \&
  {Veilleux}}]{Mazzarella:2012}
{Mazzarella}, J.~M., {et~al.} 2012, ArXiv e-prints

\bibitem[{{McGurk} {et~al.}(2011){McGurk}, {Max}, {Rosario}, {Shields},
  {Smith}, \& {Wright}}]{McGurk:2011}
{McGurk}, R.~C., {Max}, C.~E., {Rosario}, D.~J., {Shields}, G.~A., {Smith},
  K.~L., \& {Wright}, S.~A. 2011, \apjl, 738, L2

\bibitem[{{McLure} \& {Jarvis}(2002)}]{McLure:Jarvis:2002}
{McLure}, R.~J., \& {Jarvis}, M.~J. 2002, \mnras, 337, 109

\bibitem[{{Nelson} \& {Whittle}(1996)}]{Nelson:Whittle:1996}
{Nelson}, C.~H., \& {Whittle}, M. 1996, \apj, 465, 96

\bibitem[{{Nesvadba} {et~al.}(2008){Nesvadba}, {Lehnert}, {De Breuck},
  {Gilbert}, \& {van Breugel}}]{Nesvadba:2008}
{Nesvadba}, N.~P.~H., {Lehnert}, M.~D., {De Breuck}, C., {Gilbert}, A.~M., \&
  {van Breugel}, W. 2008, \aap, 491, 407

\bibitem[{{Newman} {et~al.}(2012){Newman}, {Cooper}, {Davis}, {Faber}, {Coil},
  {Guhathakurta}, {Koo}, {Phillips}, {Conroy}, {Dutton}, {Finkbeiner}, {Gerke},
  {Rosario}, {Weiner}, {Willmer}, {Yan}, {Harker}, {Kassin}, {Konidaris},
  {Lai}, {Madgwick}, {Noeske}, {Wirth}, {Connolly}, {Kaiser}, {Kirby},
  {Lemaux}, {Lin}, {Lotz}, {Luppino}, {Marinoni}, {Matthews}, {Metevier}, \&
  {Schiavon}}]{Newman:2012}
{Newman}, J.~A., {et~al.} 2012, ArXiv e-prints

\bibitem[{{Osterbrock} \& {Ferland}(2006)}]{Osterbrock:2006}
{Osterbrock}, D.~E., \& {Ferland}, G.~J. 2006, {Astrophysics of gaseous nebulae
  and active galactic nuclei}, ed. {Osterbrock, D.~E.~\& Ferland, G.~J.}

\bibitem[{{Piconcelli} {et~al.}(2010){Piconcelli}, {Vignali}, {Bianchi},
  {Mathur}, {Fiore}, {Guainazzi}, {Lanzuisi}, {Maiolino}, \&
  {Nicastro}}]{Piconcelli2010}
{Piconcelli}, E., {et~al.} 2010, ArXiv e-prints

\bibitem[{{Rosario} {et~al.}(2011){Rosario}, {McGurk}, {Max}, {Shields},
  {Smith}, \& {Ammons}}]{Rosario:2011}
{Rosario}, D.~J., {McGurk}, R.~C., {Max}, C.~E., {Shields}, G.~A., {Smith},
  K.~L., \& {Ammons}, S.~M. 2011, \apj, 739, 44

\bibitem[{{Rosario} {et~al.}(2010){Rosario}, {Shields}, {Taylor}, {Salviander},
  \& {Smith}}]{Rosario:2010}
{Rosario}, D.~J., {Shields}, G.~A., {Taylor}, G.~B., {Salviander}, S., \&
  {Smith}, K.~L. 2010, \apj, 716, 131

\bibitem[{{Sanders} {et~al.}(1988){Sanders}, {Soifer}, {Elias}, {Madore},
  {Matthews}, {Neugebauer}, \& {Scoville}}]{Sanders:1988}
{Sanders}, D.~B., {Soifer}, B.~T., {Elias}, J.~H., {Madore}, B.~F., {Matthews},
  K., {Neugebauer}, G., \& {Scoville}, N.~Z. 1988, \apj, 325, 74

\bibitem[{{Schneider} {et~al.}(2010){Schneider}, {Richards}, {Hall}, {Strauss},
  {Anderson}, {Boroson}, {Ross}, {Shen}, {Brandt}, {Fan}, {Inada}, {Jester},
  {Knapp}, {Krawczyk}, {Thakar}, {Vanden Berk}, {Voges}, {Yanny}, {York},
  {Bahcall}, {Bizyaev}, {Blanton}, {Brewington}, {Brinkmann}, {Eisenstein},
  {Frieman}, {Fukugita}, {Gray}, {Gunn}, {Hibon}, {Ivezi{\'c}}, {Kent}, {Kron},
  {Lee}, {Lupton}, {Malanushenko}, {Malanushenko}, {Oravetz}, {Pan}, {Pier},
  {Price}, {Saxe}, {Schlegel}, {Simmons}, {Snedden}, {SubbaRao}, {Szalay}, \&
  {Weinberg}}]{Schneider:2010}
{Schneider}, D.~P., {et~al.} 2010, \aj, 139, 2360

\bibitem[{{Shen} {et~al.}(2011{\natexlab{a}}){Shen}, {Liu}, {Greene}, \&
  {Strauss}}]{Shen:2011b}
{Shen}, Y., {Liu}, X., {Greene}, J.~E., \& {Strauss}, M.~A. 2011{\natexlab{a}},
  \apj, 735, 48

\bibitem[{{Shen} \& {Loeb}(2010)}]{Shen:2010a}
{Shen}, Y., \& {Loeb}, A. 2010, \apj, 725, 249

\bibitem[{{Shen} {et~al.}(2011{\natexlab{b}}){Shen}, {Richards}, {Strauss},
  {Hall}, {Schneider}, {Snedden}, {Bizyaev}, {Brewington}, {Malanushenko},
  {Malanushenko}, {Oravetz}, {Pan}, \& {Simmons}}]{Shen:2011a}
{Shen}, Y., {et~al.} 2011{\natexlab{b}}, \apjs, 194, 45

\bibitem[{{Silk} \& {Rees}(1998)}]{Silk:Rees:1998}
{Silk}, J., \& {Rees}, M.~J. 1998, \aap, 331, L1

\bibitem[{{Smith} {et~al.}(2010){Smith}, {Shields}, {Bonning}, {McMullen},
  {Rosario}, \& {Salviander}}]{Smith:2010}
{Smith}, K.~L., {Shields}, G.~A., {Bonning}, E.~W., {McMullen}, C.~C.,
  {Rosario}, D.~J., \& {Salviander}, S. 2010, \apj, 716, 866

\bibitem[{{Spoon} \& {Holt}(2009)}]{Spoon:Holt:2009}
{Spoon}, H.~W.~W., \& {Holt}, J. 2009, \apjl, 702, L42

\bibitem[{{Stoughton} {et~al.}(2002){Stoughton}, {Lupton}, {Bernardi},
  {Blanton}, {Burles}, {Castander}, {Connolly}, {Eisenstein}, {Frieman},
  {Hennessy}, {Hindsley}, {Ivezi{\'c}}, {Kent}, {Kunszt}, {Lee}, {Meiksin},
  {Munn}, {Newberg}, {Nichol}, {Nicinski}, {Pier}, {Richards}, {Richmond},
  {Schlegel}, {Smith}, {Strauss}, {SubbaRao}, {Szalay}, {Thakar}, {Tucker},
  {Vanden Berk}, {Yanny}, {Adelman}, {Anderson}, {Anderson}, {Annis},
  {Bahcall}, {Bakken}, {Bartelmann}, {Bastian}, {Bauer}, {Berman},
  {B{\"o}hringer}, {Boroski}, {Bracker}, {Briegel}, {Briggs}, {Brinkmann},
  {Brunner}, {Carey}, {Carr}, {Chen}, {Christian}, {Colestock}, {Crocker},
  {Csabai}, {Czarapata}, {Dalcanton}, {Davidsen}, {Davis}, {Dehnen},
  {Dodelson}, {Doi}, {Dombeck}, {Donahue}, {Ellman}, {Elms}, {Evans}, {Eyer},
  {Fan}, {Federwitz}, {Friedman}, {Fukugita}, {Gal}, {Gillespie}, {Glazebrook},
  {Gray}, {Grebel}, {Greenawalt}, {Greene}, {Gunn}, {de Haas}, {Haiman},
  {Haldeman}, {Hall}, {Hamabe}, {Hansen}, {Harris}, {Harris}, {Harvanek},
  {Hawley}, {Hayes}, {Heckman}, {Helmi}, {Henden}, {Hogan}, {Hogg}, {Holmgren},
  {Holtzman}, {Huang}, {Hull}, {Ichikawa}, {Ichikawa}, {Johnston}, {Kauffmann},
  {Kim}, {Kimball}, {Kinney}, {Klaene}, {Kleinman}, {Klypin}, {Knapp},
  {Korienek}, {Krolik}, {Kron}, {Krzesi{\'n}ski}, {Lamb}, {Leger},
  {Limmongkol}, {Lindenmeyer}, {Long}, {Loomis}, {Loveday}, {MacKinnon},
  {Mannery}, {Mantsch}, {Margon}, {McGehee}, {McKay}, {McLean}, {Menou},
  {Merelli}, {Mo}, {Monet}, {Nakamura}, {Narayanan}, {Nash}, {Neilsen},
  {Newman}, {Nitta}, {Odenkirchen}, {Okada}, {Okamura}, {Ostriker}, {Owen},
  {Pauls}, {Peoples}, {Peterson}, {Petravick}, {Pope}, {Pordes}, {Postman},
  {Prosapio}, {Quinn}, {Rechenmacher}, {Rivetta}, {Rix}, {Rockosi}, {Rosner},
  {Ruthmansdorfer}, {Sandford}, {Schneider}, {Scranton}, {Sekiguchi}, {Sergey},
  {Sheth}, {Shimasaku}, {Smee}, {Snedden}, {Stebbins}, {Stubbs}, {Szapudi},
  {Szkody}, {Szokoly}, {Tabachnik}, {Tsvetanov}, {Uomoto}, {Vogeley}, {Voges},
  {Waddell}, {Walterbos}, {Wang}, {Watanabe}, {Weinberg}, {White}, {White},
  {Wilhite}, {Wolfe}, {Yasuda}, {York}, {Zehavi}, \& {Zheng}}]{Stoughton:2002}
{Stoughton}, C., {et~al.} 2002, \aj, 123, 485

\bibitem[{{Strateva} {et~al.}(2003){Strateva}, {Strauss}, {Hao}, {Schlegel},
  {Hall}, {Gunn}, {Li}, {Ivezi{\'c}}, {Richards}, {Zakamska}, {Voges},
  {Anderson}, {Lupton}, {Schneider}, {Brinkmann}, \& {Nichol}}]{strateva03}
{Strateva}, I.~V., {et~al.} 2003, AJ, 126, 1720

\bibitem[{{Sturm} {et~al.}(2002){Sturm}, {Lutz}, {Verma}, {Netzer},
  {Sternberg}, {Moorwood}, {Oliva}, \& {Genzel}}]{Sturm:2002}
{Sturm}, E., {Lutz}, D., {Verma}, A., {Netzer}, H., {Sternberg}, A.,
  {Moorwood}, A.~F.~M., {Oliva}, E., \& {Genzel}, R. 2002, \aap, 393, 821

\bibitem[{{Tadhunter} {et~al.}(1988){Tadhunter}, {Fosbury}, {di Serego
  Alighieri}, {Bland}, {Danziger}, {Goss}, {McAdam}, \&
  {Snijders}}]{Tadhunter:1988}
{Tadhunter}, C.~N., {Fosbury}, R.~A.~E., {di Serego Alighieri}, S., {Bland},
  J., {Danziger}, I.~J., {Goss}, W.~M., {McAdam}, W.~B., \& {Snijders},
  M.~A.~J. 1988, \mnras, 235, 403

\bibitem[{{Treister} {et~al.}(2010){Treister}, {Natarajan}, {Sanders}, {Urry},
  {Schawinski}, \& {Kartaltepe}}]{Treister:2010}
{Treister}, E., {Natarajan}, P., {Sanders}, D.~B., {Urry}, C.~M., {Schawinski},
  K., \& {Kartaltepe}, J. 2010, Science, 328, 600

\bibitem[{{Treister} {et~al.}(2012){Treister}, {Schawinski}, {Urry}, \&
  {Simmons}}]{Treister:2012}
{Treister}, E., {Schawinski}, K., {Urry}, C.~M., \& {Simmons}, B.~D. 2012,
  \apjl, 758, L39

\bibitem[{{Tsuzuki} {et~al.}(2006){Tsuzuki}, {Kawara}, {Yoshii}, {Oyabu},
  {Tanab{\'e}}, \& {Matsuoka}}]{Tsuzuki:2006}
{Tsuzuki}, Y., {Kawara}, K., {Yoshii}, Y., {Oyabu}, S., {Tanab{\'e}}, T., \&
  {Matsuoka}, Y. 2006, \apj, 650, 57

\bibitem[{{Van Wassenhove} {et~al.}(2012){Van Wassenhove}, {Volonteri},
  {Mayer}, {Dotti}, {Bellovary}, \& {Callegari}}]{Van_Wassenhove:2012}
{Van Wassenhove}, S., {Volonteri}, M., {Mayer}, L., {Dotti}, M., {Bellovary},
  J., \& {Callegari}, S. 2012, \apjl, 748, L7

\bibitem[{{Vanden Berk} {et~al.}(2001){Vanden Berk}, {Richards}, {Bauer},
  {Strauss}, {Schneider}, {Heckman}, {York}, {Hall}, {Fan}, {Knapp},
  {Anderson}, {Annis}, {Bahcall}, {Bernardi}, {Briggs}, {Brinkmann}, {Brunner},
  {Burles}, {Carey}, {Castander}, {Connolly}, {Crocker}, {Csabai}, {Doi},
  {Finkbeiner}, {Friedman}, {Frieman}, {Fukugita}, {Gunn}, {Hennessy},
  {Ivezi{\'c}}, {Kent}, {Kunszt}, {Lamb}, {Leger}, {Long}, {Loveday}, {Lupton},
  {Meiksin}, {Merelli}, {Munn}, {Newberg}, {Newcomb}, {Nichol}, {Owen}, {Pier},
  {Pope}, {Rockosi}, {Schlegel}, {Siegmund}, {Smee}, {Snir}, {Stoughton},
  {Stubbs}, {SubbaRao}, {Szalay}, {Szokoly}, {Tremonti}, {Uomoto}, {Waddell},
  {Yanny}, \& {Zheng}}]{Berk:2001}
{Vanden Berk}, D.~E., {et~al.} 2001, \aj, 122, 549

\bibitem[{{Veilleux}(1991)}]{Veilleux:1991}
{Veilleux}, S. 1991, \apj, 369, 331

\bibitem[{{Veilleux} {et~al.}(2001){Veilleux}, {Shopbell}, \&
  {Miller}}]{Veilleux:2001}
{Veilleux}, S., {Shopbell}, P.~L., \& {Miller}, S.~T. 2001, \aj, 121, 198

\bibitem[{{V{\'e}ron-Cetty} {et~al.}(2004){V{\'e}ron-Cetty}, {Joly}, \&
  {V{\'e}ron}}]{veron_cetty:2004}
{V{\'e}ron-Cetty}, M.-P., {Joly}, M., \& {V{\'e}ron}, P. 2004, \aap, 417, 515

\bibitem[{{Wang} {et~al.}(2009){Wang}, {Chen}, {Hu}, {Mao}, {Zhang}, \&
  {Bian}}]{Wang2009}
{Wang}, J., {Chen}, Y., {Hu}, C., {Mao}, W., {Zhang}, S., \& {Bian}, W. 2009,
  \apjl, 705, L76

\bibitem[{{Wang} \& {Zhou}(2012)}]{Wang:2012}
{Wang}, X.-W., \& {Zhou}, H.-Y. 2012, \apj, 757, 124

\bibitem[{{Whittle}(1992)}]{Whittle:1992}
{Whittle}, M. 1992, \apj, 387, 109

\bibitem[{{Xu} \& {Komossa}(2009)}]{XK09}
{Xu}, D., \& {Komossa}, S. 2009, ApJ, 705, L20

\bibitem[{{Yu} {et~al.}(2011){Yu}, {Lu}, {Mohayaee}, \& {Colin}}]{Yu:2011}
{Yu}, Q., {Lu}, Y., {Mohayaee}, R., \& {Colin}, J. 2011, \apj, 738, 92

\bibitem[{{Zamanov} {et~al.}(2002){Zamanov}, {Marziani}, {Sulentic}, {Calvani},
  {Dultzin-Hacyan}, \& {Bachev}}]{Zamanov:2002}
{Zamanov}, R., {Marziani}, P., {Sulentic}, J.~W., {Calvani}, M.,
  {Dultzin-Hacyan}, D., \& {Bachev}, R. 2002, \apjl, 576, L9

\bibitem[{{Zheng} {et~al.}(1990){Zheng}, {Sulentic}, \& {Binette}}]{Zheng90}
{Zheng}, W., {Sulentic}, J.~W., \& {Binette}, L. 1990, ApJ, 365, 115

\end{thebibliography}
\end{document}